\documentclass[a4paper]{JHEP3}
\usepackage[centertags]{amsmath}
\usepackage{amssymb,amstext}
\usepackage{graphicx}
\usepackage{epsfig}
\usepackage[stable]{footmisc}
\usepackage{appendix}
\usepackage{dsfont}

\newcommand{\half}{\frac{1}{2}}
\newcommand{\threehalf}{\frac{3}{2}}

\newcommand{\e}{\epsilon}

\newcommand{\del}{\partial}

\newcommand{\sig}{{\sigma}}
\newcommand{\A}{{\alpha}}
\newcommand{\B}{{\beta}}

\newcommand{\D}{{\delta}}
\newcommand{\G}{{\gamma}}

\def\be{\begin{equation}}
\def\ee{\end{equation}}
\def\bs{\begin{split}}
\def\es{\end{split}}
\def\bea{\begin{eqnarray}}
\def\eea{\end{eqnarray}}
\def\be{\begin{equation}}
\def\ee{\end{equation}}
\def\bs{\begin{split}}
\def\es{\end{split}}
\def\bea{\begin{eqnarray}}
\def\eea{\end{eqnarray}}
\def\ie{\begin{equation}\begin{aligned}}
\def\fe{\end{aligned}\end{equation}}
\title{Superspace formulation and correlation functions of $3d$ superconformal field theories}

\author{Amin A. Nizami$^{1,a}$\,, Tarun Sharma$^{2,b}$\,, V. Umesh$^{3,b}$ \\

$^a$ DAMTP, Centre for Mathematical Sciences, Wilberforce Road, \\ 
\hspace{2mm} Cambridge CB3 0WA, UK. \\
$^b$ Department of Theoretical Physics, Tata Institute of Fundamental Research, \\
\hspace{2mm} Homi Bhabha Road, Colaba-400005, India. \\

$^1$A.A.Nizami@damtp.cam.ac.uk \\
$^2$tarun@theory.tifr.res.in \\
$^3$umesh@theory.tifr.res.in }

\abstract{We study $3d$ SCFTs in the superspace formalism and discuss superfields 
and on-shell higher spin current multiplets in free $3d$ SCFTs with $\mathcal{N}= 1,2,3,4$ and $6$ 
superconformal symmetry. For $\mathcal{N}=1$ 3d SCFTs we determine the superconformal 
invariants in superspace needed for constructing 3-point functions of higher spin operators, find the non-linear 
relations between the invariants and consequently write down all the independent 
invariant structures, both parity even and odd, for various 3-point functions of 
higher spin operators.
}

\preprint{TIFR/TH/13-25\\ DAMTP-2013-47}
\begin{document}

\section{Introduction} 

It has recently been realized that non-abelian Chern-Simons theories coupled
to fundamental matter fields in 3 dimensions are exactly solvable in the 
large $N$ limit. These theories have an interesting 'current algebra'
structure involving almost conserved higher spin fields, appear to enjoy 
invariance under nontrivial level-rank type strong-weak coupling dualities, and 
also appear to admit a bulk dual description in terms of Vasiliev's 
equations for higher spin fields \cite{Aharony:2011jz,Giombi:2011kc,Maldacena:2011jn,
Maldacena:2012sf,Banerjee:2012gh,Chang:2012kt,Aharony:2012nh,Jain:2012qi,Giombi:2012ms,
Banerjee:2012aj,Yokoyama:2012fa,GurAri:2012is,Aharony:2012ns,Jain:2013py,Takimi:2013zca,
Jain:2013gza}.

The new results obtained for large $N$ vector Chern-Simons theories are 
exciting partly because they apply to non-supersymmetric theories. 
Most of the results obtained in \cite{Giombi:2011rz,Aharony:2012nh}, 
however, have simple extensions to the supersymmetric counterparts 
of the theories studied there (see e.g. \cite{Jain:2012qi}). For instance, 
it should be possible to extend the results of Maldacena and Zhiboedov 
\cite{Maldacena:2011jn,Maldacena:2012sf} to obtain the exact form of 
the higher spin current algebra 
and the correlation functions of supersymmetric Chern-Simons coupled to 
fundamental matter fields with minimal matter content. 

In order to extend recent results in the study of matter Chern-Simons 
theories to their supersymmetric counterparts,  it would be convenient 
to have a formulation of these theories in superspace. 
Offshell superspace formulations of theories with extended 
supersymmetry are complicated and very messy. Moreover the abstract study 
of supersymmetric matter Chern Simons theories, along the lines of 
\cite{Maldacena:2011jn,Maldacena:2012sf}, does not need an offshell 
formalism. In this paper we initiate the development of 
onshell superspace techniques to study 
superconformal field theories. In particular we present a detailed 
study of free superconformal field theories in superspace using 
onshell techniques. We present a superspace construction of higher 
spin supercurrents in free theories, and describe the structural 
form of the current algebra of the corresponding higher spin currents 
once we include the effect of interactions. We also study the correlation 
functions of higher spin currents in superspace; in particular we 
conjecture that superconformal invariance and current conservation constrains the form of 
the three point functions of higher spin currents to a linear combination 
of the unique `free' structure and a parity odd structure; we present evidence 
in favor of this conjecture. 

This paper is structured as follows. In section 2 we consider 3d superspace,
and the differential form of various operators which act in it. The
construction of superconformally covariant structures in superspace is 
reviewed. Section 3 deals with specifics of $\mathcal{N}=1,\,2,\,3,\,4$ and \,$6$
superconformal symmetry in superspace and the construction of superfield
multiplets. In section 4 on-shell supercurrent multiplets for higher
spin currents in the free theory are constructed out of the superfields. 
In section 5 we make a few remarks about the structure of anomalous 
conservation equations for 3d CFTs and SCFTs with weakly broken higher 
spin symmetry. In sections 6 and 7, which are essentially independent 
of sections 3, 4 and 5 and can be read independently, we turn to 
correlation functions of $\mathcal{N}=1$ 3d SCFTs. In section 6 we give the form of the 2-point function
of a spin $s$ operator and give an elementary derivation, on the basis of symmetry and dimensional arguments, 
of the 2-point function of two spin half operators and explicitly 
compute a 2-point correlator in the free theory. In section 7 we turn to 3-point 
correlation functions - we first construct parity
even and odd superconformal invariants in superspace, determine the
myriad non-linear relations between them and then use these results
(in section 7.3) to determine the independent invariant structures
which can arise in various 3-point functions of higher spin operators. This 
section is essentially an extension, to the superconformal case,
of many of the results of \cite{Giombi:2011rz}. We build the invariants using the superconformal covariant structures constructed 
by J-H Park and H. Osborn \cite{Park:1997bq,Osborn:1998qu,Park:1998nra,Park:1999cw,Park:1999pd} augmented by the polarization spinor formalism used by \cite{Giombi:2011rz}. In appendix \ref{con} we list our conventions and some useful identities. In appendix \ref{scalfermcd} we give single trace conformal primary decomposition of a free $U(N)$ theory of a single 
complex scalar and complex fermion. In appendix \ref{repth} we present the 
full single trace superconformal primary spectrum of the theories discussed 
in section \ref{fscftss}.

\section{Superspace}\label{ss}
We begin by reviewing superspace in three dimensions and the 
covariant structures that it admits, relying heavily on the 
paper of Park \cite{Park:1999cw}. Our conventions are summarized in 
Appendix \ref{con}.

In order to study ${\cal N}=m$ superconformal field theories in 3 dimensions
we employ a superspace whose coordinates are the 3 spacetime  coordinates 
$x^\mu$ together with the $2 m$ fermionic coordinates $\theta_\alpha^a$. 
Here $\alpha= 1, 2$ is a spacetime spinor index while $a= 1 \ldots m$
is the $R$-symmetry index, where the $\theta$s (and the supercharges $Q_\alpha^a$s) are Majorana spinors that
lie in the vector representation of the $R$-symmetry group $SO({\cal N})$.
The superconformal algebra, listed in 
\eqref{N1lag} in Appendix \ref{con1}, is implemented in superspace by the 
construction
\newpage
\begin{equation}
\begin{split}
P_\mu &= -i \partial_\mu, \\
M_{\mu\nu} &= -i\left(x_{\mu}\partial_{\nu}- x_\nu \partial_\mu - 
               \frac{1}{2}\epsilon_{\mu\nu\rho}(\gamma^{\rho})_\alpha^{\mbox{ }\beta}\theta^a_\beta \frac{\partial}{\partial\theta^a_{\alpha}}\right) + {\cal M}_{\mu\nu}, \\
D &= -i\left(x^\nu\partial_\nu +\frac{1}{2}\theta^{\alpha a}\frac{\partial}{\partial\theta^\alpha_a}\right) 
      + \Delta, \\
% K_\mu &=  x^\nu M_{\nu\mu} - x_\mu D,  \\
K_\mu &= -i\left(\left(x^2 + \frac{(\theta^a\theta^a)^2}{16}\right)\partial_\mu -2 x_\mu\left(x\cdot\partial + 
          \theta^{\alpha a}\frac{\partial}{\partial\theta^\alpha_a}\right)+(\theta^a X_+ \gamma_\mu)^\beta\frac{\partial}{\partial\theta_a^{\beta}}\right) \\
 &= x^\nu M_{\nu\mu} - x_\mu D + \frac{i}2 (\theta^a\gamma_\mu X)^\alpha\frac{\partial}{\partial\theta_a^\alpha} -
        \frac{i}{16}(\theta^a\theta^a)^2\partial_\mu + \frac{(\theta^a\theta^a)}{4}(\theta^b\gamma_\mu)^\alpha\frac{\partial}{\partial\theta_b^\alpha},\\
Q_\alpha^a &= \frac{\partial}{\partial\theta^\alpha_a} - \frac{i}{2}\theta^{\beta a}(\gamma^\mu)_{\beta\alpha}\partial_\mu, \\
S_\alpha^a &= -(X_+)_\alpha^{~\beta}Q_\beta^a - i \theta^a \theta^b \frac{\partial}{\partial\theta^\alpha_b} - i \theta^{a}_\alpha \theta^{b\beta}\frac{\partial}{\partial\theta^\beta_b}
                +\frac{i}{2}(\theta^b\theta^b)\frac{\partial}{\partial\theta_a^\alpha} \\
           &= -(X_-)_\alpha^{~\beta}\frac{\partial}{\partial\theta_a^\beta} + \frac{\theta_\alpha^a}{2} D + \frac{1}{4}\epsilon_{\mu\nu\rho}(\gamma^\rho\theta^a)_\alpha M^{\mu\nu}
              -\frac{(\theta^b\theta^b)}{8}\theta^{a\beta} \del_{\beta\alpha} - \frac{3i}{4}\left(\theta_\alpha^a\theta\frac{\partial}{\partial\theta} +
               \theta^a\theta^b\frac{\partial}{\partial\theta_b^\alpha}\right), \\
I^{ab}   &= -i\left(\theta^a \frac{\partial}{\partial\theta_b} - \theta^b\frac{\partial}{\partial\theta_a}\right) + {\cal I}^{ab}~.
% -x_\alpha^{~\beta} \frac{\partial}{\partial \theta^\beta}- 
%             \frac{i}{2}\theta_\alpha x^\nu \partial_\nu- 
%             \frac{i}{4}\theta\theta\frac{\partial}{\partial\theta^\alpha}+
%             \frac{i}{2}(\gamma_\mu)_\alpha^\beta\theta_\beta \epsilon^{\mu\nu\rho}x_{\nu}\partial_{\rho}  ~.
\end{split}
\label{difalg}
\end{equation}
Here the derivative expressions act on superspace coordinates while the 
operators $\cal M$, $\Delta$ and ${\cal I}^{ab}$
act on the operators (states) which carry tensor structure, non-zero
scaling dimensions and transform non-trivially under $R$-symmetry.
All indices are contracted in matrix notation (the spinors are contracted from north-west 
to south-east, see appendix \ref{con1}) and the definitions of $X_+$, $X_-$
are given in \eqref{pmdef}. Note that  $x^2 + \frac{(\theta^a\theta^a)^2}{16} 
= \half (X_+ X_-)_\alpha^{~\alpha}$ (this combination appears in the 
expression for $K_\mu$ above). 
Below we will often have occasion to use a `supersymmetric' derivative 
operator $D^a_\alpha$ defined by 
\begin{equation}\label{susyd}
D^a_\alpha = \frac{\partial}{\partial\theta^\alpha_a} 
+ \frac{i}{2}\theta^{a\beta}\partial_{\beta\alpha}, 
\end{equation}
The operator $D^i_\alpha$ has the property that it anticommutes will all 
supersymmetry generators 
\begin{equation}\label{dqcom}
\{ D^a_\alpha, Q^b_\beta \}=0
\end{equation}
Note also that 
\begin{equation}
 \begin{split}
   \{ D^a_\alpha, D^b_\beta \} &= -P_{\alpha\beta}\delta^{ab} \\
 \end{split}
 \label{algebra3}
\end{equation}

In the sequel we will sometimes require to construct functions built out 
of coordinates in superspace that are invariant under superconformal 
transformations. Given two points in superspace, $(x_1, \theta_1)$ and 
$(x_2, \theta_2)$, it is obvious that $\theta_{12}=\theta_1-\theta_2$ 
is annihilated by the supersymmetry generators. It is also easy to verify 
that the supersymmetrized coordinate difference 
\begin{equation} \label{tx}
\tilde{x}_{12}^{\mu}=x_{12}^{\mu}+\frac{i}{2}\theta_{1}^{a\alpha}(\gamma^{\mu})_{\alpha}^{\,\,\beta}\theta^a_{2\beta}
\end{equation}
is also annihilated by all $Q_\alpha$. 

Any vector of $SO(2,1)$ may equally be regarded as a symmetrized bispinor. So
$x^\mu$ may be represented in terms of bispinors by the $2 \times 2$ matrix $X=x\cdot\gamma$. In this notation  
\eqref{tx} may be rewritten as 
\begin{equation}
(\tilde{X}_{12})_{\alpha}^{\,\,\beta}=(X_{12})_{\alpha}^{\,\,\beta}+i\theta^a_{1\alpha}\theta_{2}^{a\beta}
+\frac{i}{2}(\theta^a_{1}\theta^a_{2})\delta_{\alpha}^{~\beta}
\end{equation}

While an arbitrary function of $\theta_{12}$ and ${\tilde X}_{12}$ 
is annihilated by the supersymmetry operator, it is not, in general, 
annihilated by the generator of superconformal transformations. In order 
to build superconformally invariant expressions it is useful to note that 
\begin{equation}\label{suconf}
S^a_\alpha = I Q^a_\alpha I
\end{equation}
where $I$ is the superinversion operator, whose action on the coordinates 
of superspace is given by 
\begin{equation} \begin{split} \label{inver}
I(x^\mu)&= \frac{x^\mu}{x^2 + \frac{(\theta^a\theta^a)^2}{16}} \\
% I(\theta^a_\alpha)&= \\
\end{split}
\end{equation}
To define the superinversion properties of spinors, it is useful to define the objects 
\begin{equation}\label{pmdef}
X_{\pm}=X\pm \frac{i}{4}(\theta^a\theta^a)\mathds{1}.\end{equation}
It follows from \eqref{inver} that this object transforms 
homogeneously under inversions
\[
I(X_{\pm}) =X_{\pm}^{-1}\]
\begin{equation}
I(\theta^a_{\alpha}) = (X_{+}^{-1}\theta^a)_{\alpha}\end{equation}
\[
I(\theta^{a\beta})=-(\theta^a X_{-}^{-1})^{\beta}\] 
(Here $X$ is the $2\times 2$ matrix corresponding to a particular superspace point, not a coordinate 
difference).

Using these rules it follows that  the following objects (see e.g. 
\cite{Park:1997bq,Osborn:1998qu,Park:1998nra,Park:1999cw,Park:1999pd} 
transform homogeneously under inversions: 

\begin{equation}
(X_{ij+})_{\alpha}^{\,\,\beta}=(X_{i+})_{\alpha}^{\,\,\beta}-(X_{j-})_{\alpha}^{\,\,\beta}+i\theta_{i\alpha}^{a}\theta_{j}^{\,\beta a}\end{equation}
\begin{equation}
(X_{ij-})_{\alpha}^{\,\,\beta}=(X_{i-})_{\alpha}^{\,\,\beta}-(X_{j+})_{\alpha}^{\,\,\beta}-i\theta_{j\alpha}^{a}\theta_{i}^{\,\beta a}\end{equation}
For example, 
\begin{equation}
I \left(X_{ij+} \right)_{\alpha}^{\,\,\beta}=I \left( (X_{i+})_{\alpha}^{\,\,\beta}-(X_{j-})_{\alpha}^{\,\,\beta}+i\theta^a_{i\alpha}\theta_{j}^{a\beta}
\right) = -(X^{-1}_{i+})_\alpha^{~\gamma} (X_{ij+})_\gamma^{~\delta}(X^{-1}_{j-})_\delta^{~\beta}  \end{equation}
Moreover it may be demonstrated  \cite{Park:1997bq,Osborn:1998qu,Park:1998nra,Park:1999cw,Park:1999pd} that 
\begin{equation}
X_{ij\pm}=\tilde{X}_{ij}\pm \frac{i}4\theta_{ij}^{2}\mathds{1}
\end{equation} 
In other words $X_{ij\pm}$ transform homogeneously under inversions 
and are also annihilated by the generators of 
supersymmetry. In performing various manipulations it is useful to note that 
\begin{equation} X_{+}X_{-}=(x^{2}+\frac{1}{16}(\theta^{a}\theta^{a})^{2})\mathds{1} \end{equation}
 \begin{equation}
X_{ij+}X_{ij-}=(\tilde{x}_{ij}^{2}+\frac{1}{16}(\theta_{ij}^{a}\theta_{ij}^{a})^{2})\mathds{1}\end{equation}
so that 
\begin{equation}\label{inverse}
\begin{split}
 (X_{\pm})^{-1} = \frac{X_{\mp}}{x^2 + \frac{1}{16}(\theta^a\theta^a)^2} \\
 (X_{ij\pm})^{-1} = \frac{X_{ij\mp}}{\tilde{x}_{ij}^2 +\frac{1}{16}(\theta_{ij}^{a}\theta_{ij}^{a})^{2}}
\end{split}
\end{equation}
(note that the the $R$-symmetry index $a$ is summed over but that, throughout, $i,j~(=1,2,3)$ 
label points in superspace and are not summed over).

There also exist fermionic covariant structures
(which are identically zero in the non-supersymmetric case) which are
constructed out of the superspace co-ordinates as follows \cite{Park:1997bq,Osborn:1998qu,Park:1998nra,Park:1999cw,Park:1999pd}:
\begin{equation}
\Theta_{1\alpha}^{a}=\left((X_{21+}^{-1}\theta_{21}^a)_{\alpha}-(X_{31+}^{-1}\theta_{31}^a)_{\alpha}\right)
\end{equation}
$\Theta_{2},\,\Theta_{3}$ are defined similarly. Its transformation properties under superinversion are
\begin{equation}
\Theta_{i\alpha}^{a}\rightarrow-(X_{i-})_{\alpha}^{\,\,\beta}\Theta_{i\beta}^{b}I_{b}^{\,\,a}\,\,\,\,\,\,\,\,\Theta_{i}^{\alpha\,a}\rightarrow
                            I^{Ta}_{\,\,\,\,\,\,b}\,\Theta_{i}^{\beta\,b}(X_{i+})_{\beta}^{\,\,\alpha}
\end{equation}

The basic covariant structures $X_{ij\pm},\,\,\Theta_{i\alpha}^{a}$ are annihilated by the generators of 
supersymmetry. For this reason they form the basic 
building blocks for the construction of superconformal invariants, as we will explain in a later section. 

{\it Polarization spinors}: Since we will be dealing extensively with higher spin operators and their 
correlators, it will be useful to adopt a formalism, developed in 
\cite{Giombi:2011rz}\footnote{see also \cite{Costa:2011mg} for a similar approach}, in which the 
information about the tensor structure is encoded in {\it polarization spinors}: $\lambda_{\alpha}$. 
These auxiliary objects are book-keeping devices to keep track of the tensorial nature of correlators 
in an efficient manner. They are defined to be real, bosonic, two-component objects transforming as 
spinors of the $3d$ Lorentz group (see  \cite{Giombi:2011rz}). Being spinors in 2+1 dimensions fixes 
their transformation law under superinversions:
\begin{equation} 
\lambda_{\alpha}\rightarrow(X_{+}^{-1}\lambda)_{\alpha}\,\,\,\,\,,\,\,\,\,\,\,
\lambda^{\beta}\rightarrow-(\lambda X_{-}^{-1})^{\beta}  
\end{equation}
(This is the same as the transformation law of the $\theta$'s). 

A higher spin primary operator $J_{\mu_{1}\mu_{2}.....\mu_{s_{i}}}$ with spin $s_{i}$ can be represented in spinor components by  $J_{\alpha_{1}\alpha_{2}.....\alpha_{2s_{i}}}\equiv (\sigma^{\mu_{1}})_{\alpha_{1}\alpha_{2}}(\sigma^{\mu_{2}})_{\alpha_{3}\alpha_{4}}....(\sigma^{\mu_{s}})_{\alpha_{2s-1}\alpha_{2s}} J_{\mu_{1}\mu_{2}.....\mu_{s_{i}}}$. We note that this represents an operator supermultiplet  in contradistinction to  \cite{Giombi:2011rz} where the non-supersymmetric conformal case was considered (also, $J$ need not necessarily be a conserved current). We then define
$J_{s_{i}}\equiv\lambda^{\alpha_{1}}\lambda^{\alpha_{2}}...\lambda^{\alpha_{2s_{i}}}J_{\alpha_{1}\alpha_{2}.....\alpha_{2s_{i}}}$.

The 3-point function  $\langle J_{s_{1}}(x_{1},\theta_{1},\lambda_{1})J_{s_{2}}(x_{2},\theta_{2},\lambda_{2})J_{s_{3}}(x_{3},\theta_{3},\lambda_{3})\rangle$ is then a superconformal invariant constructed out of three points in (augmented) superspace with co-ordinates labeled by $(x_{i},\theta_{i},\lambda_{i})$. The tensor structure of the correlator, instead of being represented by indices, is encoded by the polynomial in $\lambda$'s (the 3-point function being a multinomial with degree $\lambda_{1}^{2s_{1}}\lambda_{2}^{2s_{2}}\lambda_{3}^{2s_{3}}$ for each term).

\section{Free superconformal theories in superspace\footnote{In this paper we deal exclusively
with onshell superspace. For offshell 3d superspace and multiplets 
in theories with and without gravity, see \cite{Kuzenko:2010rp, Kuzenko:2011xg}.}}\label{fscftss}

In this section we study free superconformal theories, with 
${\cal N}=1, 2, 3, 4$ and $6$ supersymmetry in superspace. \footnote{Sections 3, 4 and 5 have been worked out in collaboration with Shiraz Minwalla.}

\subsection{${\cal N}=1$}\label{supsp2}

${\cal N}=1$ superspace consists of points $z^A = (x^\mu, \theta_\alpha)$, 
where $\theta_\alpha$ is a Majorana spinor. There are two real supercharges 
$Q_\alpha$; these operators are implemented in superspace by the expressions 
\eqref{difalg} with ${\cal N}=1$.

The `minimal' free ${\cal N}=1$ theory consists of a single complex 
scalar field together with a single complex fermion. These fields are 
packaged together into a single complex ${\cal N}=1$ superfield $\Phi$ 
subject to the supersymmetric equation of motion 
\begin{equation}\label{phieq}
D^\alpha D_\alpha\Phi = 0
\end{equation}

Note that $\Phi$, like any scalar ${\cal N}=1$ superfield, may be 
expanded in components as 
\begin{equation}
\begin{split}
 \Phi &= \phi + \theta\psi + \frac{\theta^2}{2} F \\
 \bar{\Phi} &= \bar{\phi} - \theta\psi^* - \frac{\theta^2}{2} \bar{F}.
\end{split}
 \label{freef}
\end{equation}
By expanding \eqref{phieq} in components it is not  difficult to verify that 
\eqref{phieq} implies that  
$$F=0, ~~~ \partial^2 \phi=0, ~~~p_\mu \gamma^\mu \psi =0.$$
It follows that the superfield $\Phi$ subject to the equation of motion 
\eqref{phieq} actually describes a free massless scalar and fermion. 

In the case of ${\cal N}=1$ supersymmetry it is, of course, not difficult to 
find a manifestly supersymmetric offshell description of the theory. The 
equation of motion \eqref{phieq} follows by extremizing the action 
\begin{equation}\label{noa}
S= \int d^2 \theta d^3 x D_\alpha {\bar \Phi} D^{\alpha} \Phi
\end{equation}
w.r.t. $\Phi$. One way of adding interactions to the system \eqref{noa}
is to add a `superpotential' term $(\int d^2\theta W(\Phi))$ to the action; 
however we will not investigate offshell superspace in this paper.

\subsection{${\cal N}=2$}\label{supsp2}

In this case the fermionic coordinates of superspace consist of two 
copies of the minimal ${\cal N}=1$ Majorana spinor which can be labeled as
$\theta^i_\alpha$ ($i=1,2$). It is sometimes useful to group these 
coordinates into the complex pairs 
$$\theta_\alpha = \frac{1}{\sqrt{2}}(\theta^1_\alpha + i\theta^2_\alpha), ~~~
\bar{\theta}_\alpha = \frac{1}{\sqrt{2}}(\theta^1_\alpha - i \theta^2_\alpha).$$ 

In a similar manner there are two natural choices for a basis in the space 
of supersymmetries. One natural choice is to work with the supersymmetry 
operators defined in \eqref{difalg}. The commutation relations of the 
supersymmetries (and associated supersymmetric derivatives) is given by 
\begin{equation}
 \begin{split}
 \{ Q^i_\alpha, Q^j_\beta \} &= P_{\alpha\beta}\delta^{ij} \\
 \{ D^i_\alpha, D^j_\beta \} &= -P_{\alpha\beta}\delta^{ij} \\
  \end{split}
 \label{algebra2}
\end{equation}
Another choice is to work with complex supersymmetries; if we define
$$Q_\alpha = \frac{1}{\sqrt{2}}(Q^1_\alpha - i Q^2_\alpha), ~~~
D_\alpha = \frac{1}{\sqrt{2}}(D^1_\alpha - i D^2_\alpha)$$
we have 
\begin{equation}
 \begin{split}
 \{ Q_\alpha, \bar{Q}_\beta \} &= P_{\alpha\beta} \\
 \{ D_\alpha, \bar{D}_\beta \} &= -P_{\alpha\beta} \\
  \end{split}
 \label{algebra2_1}
\end{equation}
(also $\{ Q_\alpha,Q_\beta \} = \{ D_\alpha, D_\beta \} = 0$). 
In this basis the supersymmetry operators and supercovariant 
derivatives are most naturally written in terms of the complex variables 
$\theta_\alpha$; in particular for supercovariant derivatives we have 
\begin{equation}
 \begin{split}
  D_\alpha &= \frac{\partial}{\partial\theta^\alpha} + \frac{i}{2}\bar{\theta}^\beta\partial_{\beta\alpha} \\
  \bar{D}_\alpha &= \frac{\partial}{\partial\bar{\theta}^\alpha} + \frac{i}{2}\theta^\beta\partial_{\beta\alpha}\\
 \end{split}
\label{covD_supsp}
\end{equation}

It is sometimes useful to utilize `chiral' and anti chiral coordinates 
$(y_R,\theta_\alpha)$, $(y_L, \bar{\theta}_\alpha)$ where 
$$y_R^\mu = x^\mu - \frac{i}{2}\theta\gamma^\mu\bar{\theta}, 
~~~y_L^\mu =x^\mu + \frac{i}{2}\theta\gamma^\mu\bar{\theta}$$
These coordinates are useful because 
$$\bar{D}_\alpha y_R =0, ~~~D_\alpha  y_L = 0 $$
It is easily verified that 
\begin{equation}
 \begin{split}
  D_\alpha &= \frac{\partial}{\partial\theta^\alpha} + i\bar{\theta}^\beta\partial^{y_R}_{\beta\alpha} \\
  \bar{D}_\alpha &= \frac{\partial}{\partial\bar{\theta}^\alpha}\\
 \end{split}
\label{covD_supsp1}
\end{equation}
Analogous expressions may also be obtained if we choose $y_L, \theta, {\bar \theta}$ as our coordinates.

${\cal N}=2$ theories posses a $U(1)$ $R$-symmetry under which we can assign 
charges to operators. We normalize this symmetry by assigning the charges $1$ 
and $-1$ to $\theta$ and ${\bar \theta}$ respectively. It follows that the 
operators $D_\alpha$ and ${\bar D}_{\alpha}$ respectively have charges $-1$ 
and $+1$ under $R$-symmetry. Below we will sometimes use the notation 
$D \leftrightarrow D^{--}$ and ${\bar D} \leftrightarrow D^{++}$, notation 
that emphasizes these charge assignments.

The minimal free ${\cal N}=2$ theory has the same field content as the minimal 
${\cal N}=1$ theory, i.e., the propagating degrees of freedom are a single complex 
scalar and complex fermion. The manifestly supersymmetric form of these equations 
of motion is given as follows. The basic dynamical superfield $\Phi$ is required 
to be chiral
\begin{equation}
\begin{split}
\bar{D}_\alpha\Phi = D_\alpha\bar{\Phi} = 0 
\end{split}
 \label{constrain1}
\end{equation}
In addition it is required to obey the equations of motion (of a free theory):
\begin{equation}
D^\alpha D_\alpha\Phi = \bar{D}^\alpha\bar{D}_\alpha\bar{\Phi} = 0
 \label{freeeq1}
\end{equation}
These equations are solved by 
\begin{equation} 
\begin{split}
 \Phi &= \phi(y_R) + \theta\psi(y_R)=\phi + \theta\psi  -\frac{i}{2}\theta\gamma^\mu\bar{\theta}\partial_\mu\phi \\
 \end{split} 
 \label{freef2}
\end{equation}
and its complex conjugate (an anti-chiral field) is
\begin{equation}
\bar{\Phi} =\bar{\phi}(y_L) - \theta\psi^*(y_L) = \bar{\phi} -\bar{\theta}\psi^*  +\frac{i}{2}\theta\gamma^\mu\bar{\theta}\partial_\mu\bar{\phi}
\end{equation}
where $\phi$ and $\psi$ obey the free Klein Gordon and Dirac equations respectively (here  
$\theta\gamma^\mu\bar{\theta} = \theta^\alpha(\gamma^\mu)_\alpha^{~\beta}\bar{\theta}_\beta$).

As the field component of the minimal ${\cal N}=2$ theory is the same as that of the ${\cal N}=1$ theory, 
it is possible to write the ${\cal N}=2$ superfield in terms of the ${\cal N}=1$ superfield; explicitly
\begin{equation}
 \begin{split}
   \Phi_{{\cal N}=2} &= \Phi_{{\cal N}=1} + i \theta^{(2)}D^{(1)}\Phi_{{\cal N}=1} \\
   \bar{\Phi}_{{\cal N}=2} &= \bar{\Phi}_{{\cal N}=1} - i \theta^{(2)}D^{(1)}\bar{\Phi}_{{\cal N}=1} \\
 \end{split}
 \label{breakn1}
\end{equation}
(here  $\theta^{(2)}$ is the second Majorana Grassmann co-ordinate - the coordinate that belongs to ${\cal N}=2$ but not to ${\cal N}=1$ superspace - 
and the $\Phi_{{\cal N}=1}$ field has the usual expansion in the $\theta^{(1)}$ Grassmann co-ordinate.

\subsection{${\cal N}=3$}\label{supsp3}
The fermionic coordinates of superspace consist of three Majorana spinors,  $\theta^a_\alpha$ in this case. 
The indices $a$ transform in the vector representation of the $SO(3)$ $R$-symmetry. It is sometimes useful to regard vectors of the $SO(3)$ $R$-symmetry as bispinors, or 
$2 \times 2$ matrices. Vectors are easily converted to matrices by dotting their components with the 
Pauli matrices $(\sigma^a)_i\mbox{}^j$.

The field content of the minimal ${\cal N}=3$ free theory consists of two free complex scalars and two free 
complex fermions. These fields may be packaged together into a doublet of complex superfields that 
transform in the spin-$\frac{1}{2}$ of the $R$-symmetry group. The free theory is a trivial example of 
a superconformal field theory. Primary operators in any superconformal field theory are labeled by 
 $(\Delta,j,h)$ where $\Delta$ is the scaling dimension, $j$ is the spin and $h$ is the `$R$-symmetry spin' (i.e. the 
quantum number that describes the $R$-symmetry representation of the primary operator). In this notation the 
free superfield described above transforms in the representation $(\frac{1}{2},0,\frac{1}{2})$. 
The doublet of free superfields obey the `equations of motion'
\begin{equation}
 D_\alpha^{\{ij}\Phi^{k\}} = 0,
 \label{con3}
\end{equation}
This equation of motion has a simple interpretation; it follows from the analysis of 
unitary representations of the superconformal algebra that a representation with 
quantum numbers $(\frac{1}{2},0,\frac{1}{2})$ has a null state with quantum numbers 
$(1,\frac{1}{2},\frac{3}{2})$; the equation \eqref{con3} is simply the assertion that this
null state vanishes. 

The equations of motion \eqref{con3} may be analyzed as follows. Let us denote the two components of 
the doublet superfield $\Phi$ by 
$\Phi^+$ and (superscripts denote $R$-symmetry charge; a single $+$ denotes charge $\frac{1}{2}$). 
The equations of motion assert that 
\begin{equation}
 \begin{split}
  2 D^{(3)}_\alpha\Phi^+ - \sqrt{2}D^{++}_\alpha\Phi^- &= 0 \\
  2 D^{(3)}_\alpha\Phi^- + \sqrt{2}D^{--}_\alpha\Phi^+ &= 0 \\
 \end{split}
\label{constraint3}
\end{equation}
It is possible to solve for $\Phi^+$ and $\Phi^-$ in terms of a single ${\cal N}=2$ chiral superfield $\varphi^+$ and 
a single antichiral superfield $\varphi^-$; we find 
\begin{equation}
 \begin{split}
   \Phi^+ &= \varphi^+ + \frac{1}{\sqrt{2}} \theta^{(3)}D^{++}\varphi^- \\
   \Phi^- &= \varphi^- - \frac{1}{\sqrt{2}} \theta^{(3)}D^{--}\varphi^+ \\
 \end{split}
 \label{constsol}
\end{equation}
These ${\cal N}=2$ superfields in turn obey the free ${\cal N}=2$ equations of motion
\begin{equation}
D^{\alpha} D_\alpha\varphi^+ = \bar{D}^\alpha\bar{D}_\alpha\varphi^- = 0
 \label{freeeq2}
\end{equation}
demonstrating that the propagating degrees of freedom are twice that of the ${\cal N}=2$ theory. 

The final expression of the ${\cal N}=3$ superfield in terms of the component fields, after we have 
solved for the \eqref{freeeq2}, is given by 
\begin{equation}
 \Phi^k = \phi^k -\frac{1}{\sqrt{2}} \theta^{kl\alpha}\psi_{l\alpha} 
 - \frac{1}{4}\epsilon^{abc}\theta^{a\alpha}\theta^{b\beta}(\sigma^c)^{kl}\partial_{\alpha\beta}\phi_l 
 + \frac{1}{12\sqrt{2}}\epsilon^{abc}\theta^{a\alpha}\theta^{b\beta}\theta^{c\gamma}\partial_{\alpha\beta}\psi_\gamma^k
 \label{covex}
\end{equation}
In the last term the $\alpha, \beta, \gamma$ indices are completely symmetrized and $k=1,2.$
Here $a,b$ are vector $SO(3)$ indices and $i,j,k$ are spinor indices. Note that 
\eqref{covex} hold only when the component fields obey the free equations of motion.

\subsection{${\cal N}=4$}\label{supsp4}
In this case we have four Majorana spinor coordinates $\theta_\alpha^a$ 
lying in the $4$ of the $R$-symmetry group $SO(4)$.
The superfield $\Phi^{i}$ is a Weyl spinor of $SO(4)$\footnote{\label{1}See appendix \ref{con13} for $SO(4)$ conventions}. The
${\cal N}=4$ chirality constraint is
\begin{equation}\label{n4cons}
 D^{{\tilde i}\{j}_\alpha\Phi^{k\}} = D^{{\tilde i}j}_\alpha\Phi^k + D^{{\tilde i}k}_\alpha\Phi^j = 0.
\end{equation}
To understand the field content of the minimal ${\cal N}=4$ theory, we split the ${\cal N}=4$
chirality constraint into a part that constrains the $\theta_\alpha^{(4)}$ 
dependence and a part that's purely ${\cal N}=3$.  We begin by choosing an ${\cal N}=3$ subspace, which we take as the $1,2,3$
directions. The remaining $4$ direction is the orthogonal direction. A chiral (top-half) part of a $SO(4)$ Weyl spinor is the 
Dirac spinor in three dimensions. The $SO(4)$ vector $D_\alpha^{(a)}$ decomposes into an $SO(3)$ vector $D^{(a)} \mbox{~for~} a = 1,2,3$ and a scalar
$D^{(4)}_\alpha$. This can be seen as the symmetric and antisymmetric part of the matrix $D_\alpha^{i{\tilde i}}$ respectively.
The antisymmetric part contains only $D_\alpha^{(4)}$ and the symmetric part is the $D_\alpha^{ij}$ which is purely along the
1,2 and 3 directions.
When the above chirality constraint is analyzed, one finds
\begin{equation}\label{n4break}
 D^{(4)}_\alpha\Phi^k = -\frac{i}{3}D^{ki}_\alpha\Phi_i
\end{equation}
where on the L.H.S. we have the the supercovariant derivative along the $4$ 
direction in the $SO(4)$ $R$-symmetry space;
on the R.H.S. we have the symmetric part of the $D^{i{\tilde i}}_\alpha$ 
supercovariant derivative, which is purely along the (orthogonal) $SO(3)$ subspace.
All the spinor indices are now thought of as $SO(3)$ (Dirac) ${\cal N}=3$ 
spinor indices. This equation is the analog of
\eqref{constraint3} in the present case.

Solving \eqref{n4break} shows that the chiral ${\cal N}=4$ superfield $\Phi^{k}$ 
is completely determined in terms
of a single ${\cal N}=3$ chiral superfield $\varphi_i$ as
\begin{equation}\label{n4soln}
 \Phi^{k} = \varphi^{k} - \frac{i}{3}\theta^{(4)}D^{ki}\varphi_i.
\end{equation}
Thus, we see that the minimal field content of the ${\cal N}=4$ theory 
is the same as that of ${\cal N}=3$. An explicit component field expression 
can now be obtained from \eqref{n4soln} by using \eqref{covex} for the $\varphi^k$.

\subsection{${\cal N}=6$}\label{supsp6}
In this case we have six Majorana spinor coordinates $\theta_\alpha^a$ 
lying in the vector representation of the $R$-symmetry group $SO(6)~(\equiv SU(4))$.
The superfield $\Phi^I$ is a Weyl spinor of $SO(6)$\footnote{\label{2}See appendix \ref{con14} for $SO(6)$ conventions}, which is the $4$ of $SU(4).$
The field $\Phi^I$ satisfies the `chirality constraint'
\footnote{We briefly use upper case $I,J$ which take values
$1,\dots4$ for the $SU(4)$
indices in \eqref{N6chira} to avoid confusion with the lower case $i,j$ which appear in the ${\cal N}=4$
equations.}
\be\begin{split}\label{N6chira}
\quad D_\A^{IJ}\Phi^K &= D_\A^{JK}\Phi^I = D_\A^{KI}\Phi^J \\
\end{split}
\ee
To understand the field content
of the minimal ${\cal N}=6$ theory, we proceed as above and split the ${\cal N}=6$ chirality
constraint into an ${\cal N}=4$ part and another piece which describes the $\theta^{(5)}$ and $\theta^{(6)}$
dependences. We begin by choosing an ${\cal N}=4$ subspace, which we take as the $1,2,3 \mbox{~and~} 4$
directions. The remaining $5,6$ directions are the orthogonal directions. In the conventions we have chosen, it
may be checked that a Weyl spinor $\Phi^I$ of $SO(6)$ decomposes as one chiral $\Phi^i$ spinor ($i=1,2$)
and one anti-chiral (bottom half of the $SO(4)$ spinor) $\Phi^{{\tilde i}}$ (${\tilde i}=3,4$) of the $SO(4)$ sub-group
\footnote{We adopt the convention wherein the un-tilded indices $i$ take values 1,2 and the tilded indices ${\tilde i}$
take values 3,4}.
Similarly, the $SO(6)$ vector (the $(4\times4)_{\mbox{antisym}}$ of $SU(4)$) decomposes into two scalars and one $SO(4)$ vector.
In matrix language, one can construct the antisymmetric matrix $D_\alpha^{IJ}$ and observe that the two scalars ($D_\alpha^{(5)}$
and $D_\alpha^{(6)}$) form the linear combinations $i\sqrt{2}D_\alpha = D_\alpha^{(5)}-iD_\alpha^{(6)}$
and $i\sqrt{2}{\bar D}_\alpha = D_\alpha^{(5)}+iD_\alpha^{(6)}$, when $I,J = {\tilde i},{\tilde j}$ and $I,J = i,j$ respectively.
On the other hand, when $I,J = {\tilde i},j$ (or vice-verse) we get the (single) vector which involves only $D_\alpha^{(a)}$
where $a = 1,\ldots4.$
We can pick any two terms from the above equation (we choose the first and third) and analyze them as follows
\begin{equation}\label{n6chirdec}
 \begin{array}{cccc}
  D_\alpha^{ij}\Phi^k &= D_\alpha^{ki}\Phi^j  &~~~~~~~ D_\alpha^{{\tilde i}{\tilde j}}\Phi^{\tilde k} &=   D_\alpha^{{\tilde k}{\tilde i}}\Phi^{\tilde j} \\
  D_\alpha^{{\tilde i}j}\Phi^k &= D_\alpha^{k{\tilde i}}\Phi^j  &~~~~~~~ D_\alpha^{i{\tilde j}}\Phi^{{\tilde k}} &= D_\alpha^{{\tilde k}i}\Phi^{{\tilde j}}\\
  D_\alpha^{ij}\Phi^{{\tilde k}} &= D_\alpha^{{\tilde k}i}\Phi^j &~~~~~~~ D_\alpha^{{\tilde i}{\tilde j}}\Phi^k &= D_\alpha^{k{\tilde i}}\Phi^{{\tilde j}}
 \end{array}
\end{equation}
The second equation in each of the two sets above is just the ${\cal N}=4$ chirality condition \eqref{n4cons} for each of the fields
$\Phi^i$ and $\Phi^{\tilde i}$. It remains to analyze the first and third equations from each of the two sets. The first set reads
\begin{equation}
 \begin{split}
  {\bar D}_\alpha\Phi^k = 0 &~~~~~~~ D_\alpha\Phi^{{\tilde k}} = 0
 \end{split}
\end{equation}
where ${\bar D}_\alpha = \frac{1}{\sqrt{2}}(D^{(5)}_\alpha + i D^{(6)}_\alpha)$ and $D_\alpha = \frac{1}{\sqrt{2}}(D^{(5)}_\alpha - i D^{(6)}_\alpha).$
Thus, $\Phi^k$ and $\Phi^{\tilde k}$ can be thought of as two independent `chiral' and `anti-chiral' superfields and
we can accordingly expand them in the $\theta_\alpha = \frac{1}{\sqrt{2}}(\theta^{(5)}_\alpha + i \theta^{(6)}_\alpha)$
and ${\bar \theta}_\alpha = \frac{1}{\sqrt{2}}(\theta^{(5)}_\alpha - i \theta^{(6)}_\alpha)$ coordinates. Let's now analyze the third
equation from the above set. They are
\begin{equation}
 \begin{split}
  {\bar D}_\alpha\Phi^{{\tilde k}} = \frac{i}{2\sqrt{2}}D_\alpha^{{\tilde k}i}\Phi_i &~~~~~~~ D_\alpha\Phi^{k} = \frac{i}{2\sqrt{2}}D_\alpha^{k{\tilde i}}\Phi_{{\tilde i}}
 \end{split}
\end{equation}
Solving the above equations leads us to the following result for $\Phi^k$ and $\Phi^{\tilde k}$
\begin{equation}
 \begin{split}\label{n6exps}
   \Phi^k &= \varphi^k +  \frac{i}{2\sqrt{2}}\theta D^{k{\tilde i}}\varphi_{\tilde i} - \frac{i}{2}\theta\gamma^\mu{\bar\theta}\del_\mu\varphi^k \\
   \Phi^{{\tilde k}} &= \varphi^{{\tilde k}} +\frac{i}{2\sqrt{2}}\bar{\theta} D^{{\tilde k}i}\varphi_i + \frac{i}{2}\theta\sigma^\mu{\bar\theta}\del_\mu\varphi^{{\tilde k}}
 \end{split}
\end{equation}
Where $k = 1,2$ and ${\tilde k} = 3,4$ make up the full ${\cal N}=6$ multiplet, and
$\theta_\alpha = \frac{1}{\sqrt{2}}(\theta^{(5)}_\alpha + i \theta^{(6)}_\alpha)$
and ${\bar \theta}_\alpha = \frac{1}{\sqrt{2}}(\theta^{(5)}_\alpha - i \theta^{(6)}_\alpha)$.
Thus, we see that the field content of the minimal ${\cal N}=6$ theory consists of two 
independent ${\cal N}=4$ fields, $\varphi^k$ and $\varphi^{{\tilde k}}$. \eqref{covex} and 
\eqref{freeeq2} can now be used in \eqref{n6exps} to obtain explicit component field expression 
for $\Phi^K$.

\section{Currents}\label{supcur}
In this section we describe the construction of conserved currents in the theories 
discussed above. These currents constitute the full local gauge invariant operator 
spectrum of the theories considered. In the non-supersymmetric case the bosonic conserved 
currents and the violation, due to interactions, of their conservation by $\frac{1}{N}$ 
effects play a central role in the solution of three point functions in these 
theories \cite{Maldacena:2011jn,Maldacena:2012sf}. 
The currents we consider in this section are the supersymmetric extension of the bosonic currents 
considered in \cite{Giombi:2011rz,Maldacena:2011jn,Maldacena:2012sf}. We 
construct the supercurrents, using the onshell 
superspace described in sections \ref{ss} and \ref{fscftss}, in terms of onshell 
superfields and supercovariant derivatives.

\subsection{General structure of the current superfield}\label{genstrcur}
Let us start by first describing the structure of the ${\cal N}=1$ supercurrents. A general 
spin $s$ supercurrent multiplet can be written as a superfield carrying $2s$
spacetime spinor indices and can be expanded in components as follows
\begin{equation}\label{current}
\Phi^{\alpha_1\alpha_2\ldots\alpha_{2s}} = \phi^{\alpha_1\alpha_2\ldots\alpha_{2s}} + \theta_\alpha\psi^{\alpha\alpha_1\alpha_2\ldots\alpha_{2s}}
 + \theta^{\{\alpha_1}\chi^{\alpha_2\ldots\alpha_{2s}\}} + \theta^\alpha\theta_\alpha B^{\alpha_1\alpha_2\ldots\alpha_{2s}}
\end{equation}
where all the indices $\alpha_1\mbox{,}\alpha_2\mbox{,}\ldots\alpha_{2s}$ are symmetrized.
The conservation (shortening) condition for the supercurrent is
\begin{equation}\label{shrtcondt}
D_{\alpha_1}\Phi^{\alpha_1\alpha_2\ldots\alpha_{2s}} =0
\end{equation}
where $D_\A$ is the supercovariant derivative given by
\begin{equation}\label{sc}
D_\alpha = \frac{\partial}{\partial\theta^\alpha} +\frac{i}{2} \theta^\beta\partial_{\beta\alpha}
\end{equation}

Using eqs.(\ref{sc}) and (\ref{current}) we obtain
\begin{eqnarray}\label{condt}
 \delta_{\alpha_1}^{\mbox{ }\{\alpha_1}\chi^{\alpha_2\ldots\alpha_{2s}\}} 
 &+& \theta_{\alpha_1}(2B^{\alpha_1\alpha_2\ldots\alpha_{2s}} 
 -\frac{i}{2} \partial^{\mbox{ }\alpha_1}_\beta \phi^{\beta\alpha_2\ldots\alpha_{2s}}) \nonumber\\
  &-& \frac{i}{2} \theta^{2} \partial_{\alpha\alpha_1}\psi^{\mbox{ }\alpha\alpha_1\alpha_2\ldots\alpha_{2s}}
 +\frac{i}{2} \theta^\beta\partial_{\beta\alpha_1}\theta^{\{\alpha_1}\chi^{\alpha_2\ldots\alpha_{2s}\}} = 0
\end{eqnarray}
This implies
\begin{equation}\label{chicond}
 \chi^{\alpha_2\ldots\alpha_{2s}} = 0
\end{equation}
while the symmetric part of the $\theta$ component gives
\begin{equation}\label{bcond}
B^{\alpha_1\alpha_2\ldots\alpha_{2s}} = \frac{i}{4}\partial^{\mbox{ }\{\alpha_1}_{\beta}\phi^{|\beta|\alpha_2\ldots\alpha_{2s}\}}
\end{equation}
whereas the antisymmetric part gives
\begin{equation}\label{phicond}
 \epsilon_{\alpha_1\alpha_2}\partial^{\alpha_1}_{\mbox{ }\beta}\phi^{\beta\alpha_2\ldots\alpha_{2s}} = 0
 \Rightarrow \partial_{\alpha_1\alpha_2}\phi^{\alpha_1\alpha_2\ldots\alpha_{2s}} = 0
\end{equation}
which is the current conservation equation for the current $\phi$.
Since $\chi = 0$, the $\theta\theta$ component gives the current conservation equation for $\psi$
\begin{equation}\label{psicond}
 \partial_{\alpha\alpha_1}\psi^{\alpha\alpha_1\ldots\alpha_{2s}} = 0
\end{equation}
Thus the form of the supercurrent multiplet for a spin $s$ conserved current is
\begin{equation}\label{current1}
 \Phi^{\alpha_1\alpha_2\ldots\alpha_{2s}} = \phi^{\alpha_1\alpha_2\ldots\alpha_{2s}} + \theta_\alpha\psi^{\alpha\alpha_1\alpha_2\ldots\alpha_{2s}}
 +\frac{i}{4} \theta^\alpha\theta_\alpha \partial^{\{\alpha_1}_{\mbox{ }\beta}\phi^{|\beta|\alpha_2\ldots\alpha_{2s}\}}
\end{equation}

The general structure of the current superfield described above goes through for higher 
supersymmetries as well. For higher supersymmetries the conservation equation reads
\be\label{Ncurcons}
D^a_{~\A_1} \Phi^{\A_1\A_2\ldots\A_{2s}}=0
\ee
where $a=1,2\ldots {\cal N}$ is the R-symmetry index\footnote{Note that for ${\cal N} > 1$, 
\eqref{Ncurcons} is true only for R-symmetry singlet currents. For currents 
carrying non-trivial R-symmetry representation the shortening condition is different. In 
this paper we will only need the shortening condition \eqref{Ncurcons}.}.
In the case of an ${\cal N}=m$ spin-$s$ current multiplet, 
the currents $\phi^{\alpha_1\alpha_2\ldots\alpha_{2s}}$ and 
$\psi^{\alpha\alpha_1\alpha_2\ldots\alpha_{2s}}$ are themselves ${\cal N}=m-1$ 
spin $s$ and spin $s+\half$ conserved current superfields (depending 
on the grassmann coordinates $\theta_\alpha^{a}$: $a=1,\ldots m-1$)
while the $\theta_\A$ in \eqref{current1} is the left over grassmann coordinate 
$\theta_\alpha^m$. Thus we see the general structure of the supercurrent multiplets: 
An ${\cal N}=m$ spin $s$ supercurrent multiplet breaks up into two ${\cal N}=m-1$ 
supercurrents with spins $s$ and $s+\half$ respectively. 

This structure can be used to express higher supercurrents superfields in term of 
components. For instance, the ${\cal N}=2$ spin $s$ currents superfield can be 
expanded in components as follows
\be\begin{split}\label{N2currcomp}
\Phi^{\A_1\A_2\ldots\A_{2s}}&= \varphi^{\A_1\A_2\ldots\A_{2s}} + \theta^a_\A (\psi^a)^{\A\A_1\A_2\ldots\A_{2s}} 
                            + \half \e_{ab} \theta^a_\A\theta^b_\B\mathcal{A}^{\A\B\A_1\A_2\ldots\A_{2s}} \\ 
                          & \qquad + \text{term involving derivatives of $\varphi,\psi^a$ and $\mathcal{A}$}
\end{split}
\ee
where $a,b$ are R-symmetry indices and take values in $\{1,2\}$. The conformal state content so 
obtained, namely $(\varphi, \psi^1, \psi^2, \mathcal{A})$ above, match exactly with the 
decomposition of spin $s$ supercurrent multiplet into conformal multiplets presented in 
appendix \ref{N2spectrum}.

% Therefore, any ${\cal N}=m$ current superfield can be expanded in the $\theta_\alpha^m$ coordinate where the superfields in the expansion
% themselves are ${\cal N}=m-1$ spin $s$ and spin $s+\half$ conserved current superfields.

\subsection{Free field construction of currents}\label{ffc}
In this section we give explicit construction of the conserved supercurrent 
discussed in previous subsection in term of free superfields.

\subsubsection{${\cal N} =1$}\label{lsmult1}
The spin $s$ supercurrent here can be expressed in term of the ${\cal N}=1$ superfield 
$\Phi$ as follows
\begin{equation}
 J^{(s)} = \sum_{r=0}^{2s} (-1)^{\frac{r(r+1)}{2}}{2s \choose r} D^r\bar{\Phi} D^{2s-r}\Phi
 \label{expression1}
\end{equation}
where $J^{(s)} = \lambda^{\alpha_1}\lambda^{\alpha_2}\cdots\lambda^{\alpha_{2s}} J_{\alpha_1\alpha_2\cdots\alpha_{2s}}$
and $D = \lambda^\alpha D_\alpha$, and $\lambda_\alpha$s are polarization spinors and $s = 0, \half,1\ldots.$ The currents are of both integral
and half-integral spins. It can be verified that the above is the unique expression for the conserved spin-$s$ current
in ${\cal N}=1$ free field theory. In equations, the following holds
\begin{equation}
 \frac{\partial}{\partial\lambda^\alpha}D^\alpha J^{(s)} = 0.
 \label{conser}
\end{equation}
We note here that the stress tensor lies in the spin $\threehalf$ current supermultiplet (which also contains the
supersymmetry current), and thus is conserved exactly even in interacting theory.

\subsubsection{${\cal N} =2$}\label{lsmult2}
We give the expression of the conserved current in terms of the free ${\cal N}=2$ superfield $\Phi$ and
its complex conjugate ${\bar\Phi}$.
\begin{equation}
 J^{(s)} = \sum_{r=0}^{s}\left\{(-1)^{r(2r+1)} {2s \choose 2r} \partial^r \bar{\Phi} \partial^{s-r}\Phi 
           + (-1)^{(r+1)(2r+1)} {2s \choose 2r+1} \partial^r {\bar D}\bar{\Phi} \partial^{s-r-1} D\Phi \right\}
 \label{expression2_1}
\end{equation}
where $\partial = i\lambda^\alpha\gamma^\mu_{\alpha\beta}\lambda^\beta\partial_\mu$, $D = \lambda^\alpha D_\alpha$
and $s = 0,1,2\ldots$. The spin $1$ supercurrent multiplet contains the stress tensor, 
supersymmetry current and $R$-current, and its conservation holds even in the interacting 
superconformal theory. 

As described above in subsection \ref{genstrcur} these ${\cal N}=2$ currents can be 
decomposed into ${\cal N}=1$ currents. It is straightforward to check that the currents 
\ref{expression2_1} when expanded in $\theta^2_{\A}$ as in \eqref{current1} correctly 
reproduce the ${\cal N}=1$ currents \eqref{expression1}. This give a consistency check 
of these ${\cal N}=2$ currents.

\subsubsection{${\cal N} =3$}\label{lsmult3}
The ${\cal N}=3$ chirality constraint on the matter superfield $\Phi^{k}$ is 
\be\begin{split}\label{N3chir}
D^{\{ij} \Phi^{k\}} &= D^{ij} \Phi^{k}+ D^{ik} \Phi^{j}+ D^{jk} \Phi^{i}=0 \\
{\rm or~equivalently ~~~~~~} 
D^{ij} \Phi^{k} &= -\frac{1}{3} \big( D^{il} \Phi_{l} \e^{jk} + D^{jl} \Phi_{l} \e^{ik}\big) \\
\end{split}
\ee
where $D^{ij}_\A= (\sigma^a)^{ij} D^a_\A$.

From this chirality constraint the following identities, which would be useful in 
proving current conservation, can be derived\footnote{see appendix \ref{con12} for
$SO(3)$ conventions} 
\be\begin{split}\label{N3D2rel}
D^{ij}_\A D^{mn}_\B \Phi^k = \half \big( i\del_{\A\B} \Phi^i \e^{jm}\e^{nk} 
                              + i\del_{\A\B} \Phi^i \e^{jn}\e^{mk} 
                              + i\del_{\A\B} \Phi^j \e^{im}\e^{nk} 
                              + i\del_{\A\B} \Phi^j \e^{in}\e^{mk} \big)
\end{split}
\ee
Contracting various indices, the following relations can be obtained from \eqref{N3D2rel} as 
corollaries
\be\begin{split}\label{N3D2cor}
D^{\A ij} D^{mn}_\A \Phi^k &= 0 \\
D^{ij}_\A D^{mk}_\B \Phi_k &= -\frac{3}{2} \big( i\del_{\A\B} \Phi^i \e^{jm} + i\del_{\A\B} \Phi^j \e^{im} \big) \\
D^{ij}_\A D_{ij\B} \Phi^k &= -3i \del_{\A\B} \Phi^k = \frac{2}{3} D^k_j D^{ji} \Phi_i \\
\end{split}
\ee
We give here the expression for the conserved currents in terms of the ${\cal N}=3$ superfield $\Phi^i$.
\begin{equation}
\begin{split}
  J^{(s)} = \sum_{r=0}^s (-1)^r {2s \choose 2r} \partial^r\bar{\Phi}^i\partial^{s-r}\Phi_i + \frac{2}{9}\sum_{r=0}^{s-1}(-1)^{r+1}{2s \choose 2r+1}\partial^r D_i^{\mbox{ }j}\bar{\Phi}^i\partial^{s-r-1}D_j^{\mbox{ }k}\Phi_k \\
  J^{(s+\frac{1}{2})} = \sum_{r=0}^s \left\{(-1)^r {2s+1 \choose 2r} \partial^r\bar{\Phi}^i\partial^{s-r}D_i^{\mbox{ }j}\Phi_j + (-1)^{r+1}{2s+1 \choose 2r+1}\partial^r D_i^{\mbox{ }j}\bar{\Phi}^i\partial^{s-r}\Phi_j\right\}\\
\end{split} 
 \label{expression3}
\end{equation}
where $\partial = i\lambda^\alpha\gamma^\mu_{\alpha\beta}\lambda^\beta\partial_\mu$, $D = \lambda^\alpha D_\alpha$
and $s = 0,1,2\ldots$. The stress energy tensor in this case lies the spin $\half$ supercurrent multiplet 
along with the $R$-current and supersymmetry currents. The conservation of this supercurrent holds 
exactly even in the interacting superconformal theory.

\subsubsection{${\cal N} =4$}\label{lsmult4}
The $R$-symmetry in this case is $SO(4)$ (equivalently $SU(2)_l\times SU(2)_r$)\footnote{The indices ${a,b..}$ take values 
${1,2,3,4}$ and represent the vector indices of $SO(4)$ while the fundamental indices of the $SU(2)_l$ 
and $SU(2)_r$ are denoted by ${i,j...}$ and {$\tilde{i},\tilde{j}$...} }. The supercharges $Q_\A^{\tilde{i}i}$ 
transform in the $4$ of $SO(4)$(equivalently $(2,2)$ of $SU(2)_l\times SU(2)_r$). The two matter superfields 
transform in the $(2,0)$ representation which implies that the scalar transforms in the $(2,0)$
while the fermions transform in $(0,2)$. The matter multiplet again satisfies a `chirality' constraint
\begin{equation}
\begin{split}\label{N4chir}
D^{\tilde{i}\{i} \Phi^{j\}} &= D^{\tilde{i}i} \Phi^{j} + D^{\tilde{i}j} \Phi^{i}=0, \\
{\rm or ~ equivalently~~~~~} 
D^{\tilde{i}i} \Phi^{j} &= -\half \e^{ij} D_\A^{\tilde{i}k}\Phi_k.
\end{split}
\end{equation}
where $D^{\tilde{i}j}_\A = (\bar\sigma^a)^{\tilde{i}j} D^a_\A$.

From this chirality constraint the following identities, useful in 
proving  current conservation, can be derived\footnotemark[1]
\begin{equation}\label{N4D2rel}
D^{\tilde{i}i}_\A D^{\tilde{j}j}_\B \Phi^k =2i\del_{\A\B}\Phi^i \e^{\tilde{i}\tilde{j}} \e^{jk}
\end{equation}
Contracting various indices, the following equations can be obtained from \eqref{N4D2rel} as 
corollaries
\be\begin{split}\label{N4D2cor}
D^{\A\tilde{i}i} D^{\tilde{j}j}_\A \Phi^k &=0 \\
D^{\tilde{i}i}_\A D^{\tilde{j}j}_\B \Phi_j &= -4i\del_{\A\B}\Phi^i \e^{\tilde{i}\tilde{j}} \\
D_\A^{\tilde{i}j}D_{\B\tilde{i}k}\Phi^k &= 2D_\A^{\tilde{i}i}D_{\B\tilde{i}i}\Phi^j= 8i\del_{\A\B}\Phi^j.
\end{split}
\ee
Using these equations it is straightforward to show that the following currents are conserved.
\be\label{N4curr}
J^{(s)}= \sum^s_{r=0} (-1)^r {2s \choose 2r} ~\del^r\bar\Phi^i ~\del^{s-r}\Phi_i 
           + \frac{1}{8}\sum^{s-1}_{r=0} (-1)^r {2s \choose 2r+1} ~\del^r D^{\tilde{i}i} \bar\Phi_i 
           ~\del^{s-r-1}D_{\tilde{i}j} \Phi^j. \\
\ee
where $\partial = i\lambda^\alpha\gamma^\mu_{\alpha\beta}\lambda^\beta\partial_\mu$, 
$D = \lambda^\alpha D_\alpha$ and $s = 0,1,2\ldots$. In this theory the stress 
energy tensor lies in the $R$-symmetry singlet spin zero supercurrent multiplet $(1,0,\{0,0\})$.

\subsubsection{${\cal N}=6$}\label{lsmult6}
The field content of this theory is double of the field content of the ${\cal N}=4$ theory. In ${\cal N}=2$ 
language the field content is 2 chiral and 2 antichiral multiplets in fundamental of the gauge group. 
The $R$-symmetry in this theory is $SO(6)~(\equiv SU(4))$ under which the 
supercharges transform in vector representation ($6$ of $SO(6)$) while the 2+2 chiral and antichiral multiplets 
transform in chiral spinor representation ($4$ of $SU(4)$). 

The ${\cal N}=6$ shortening (chirality) condition on the matter multiplet is\footnote{Here we revert back to lower case
letters for the $SU(4)$ indices $i,j$ (taking values $1,\ldots4$) as there is no confusion with other
$R$ indices.}
\be\begin{split}\label{N6chir}
\quad D_\A^{ij}\Phi^k &= D_\A^{jk}\Phi^i = D_\A^{ki}\Phi^j \\
\text{or equivalently} \quad D^a_\A \Phi^k &= -\frac{1}{10} D^b_\A \Phi^l (\bar{\G}^{ab})_l^{~k} \\
\end{split}
\ee

From this chirality constraint the following identities, which are useful in 
proving  current conservation, can be derived\footnotemark[2] 
\be\begin{split}\label{N6D2rel}
D^a_\A D^b_\B \Phi^k &= \frac{i}{2} \del_{\A\B}\Phi^k \D^{ab} 
                         + \frac{i}{4} \del_{\A\B} \Phi^l (\bar{\gamma}^{ab})_l^{~k}, \\
\text{or equivalently} \quad 
D^{ij}_\A D^{mn}_\B \Phi^k &= - i\del_{\A\B} \left( \e^{ijmn} \Phi^k 
                              + \e^{kjmn} \Phi^i + \e^{ikmn} \Phi^j 
                              - \e^{ijkn} \Phi^m - \e^{ijmk} \Phi^n \right) \\
\end{split}\ee

Taking the complex conjugate of equations \eqref{N6chir} and \eqref{N6D2rel}, and using the 
property that $\gamma^{ab}$ and $\bar{\gamma}^{ab}$ are antihermitian, we get 
\be\begin{split}\label{N6chirc}
D_\A^{ij}\bar{\Phi}_k &= \frac{1}{3} \left( D_\A^{il}\bar{\Phi}_l \D_k^j 
                                                        - D_\A^{jl}\bar{\Phi}_l \D_k^i \right) \\
\text{or equivalently} \quad D^a_\A \bar{\Phi}_k &= \frac{1}{10} D^b_\A (\bar{\G}^{ab})_k^{~l}\bar{\Phi}_l \\
\end{split}
\ee
and 
\be\begin{split}\label{N6D2relc}
D^a_\A D^b_\B \bar{\Phi}_k &= \frac{i}{2} \del_{\A\B}\bar{\Phi}_k \D^{ab} 
                             - \frac{i}{4} \del_{\A\B} (\bar{\gamma}^{ab})_k^{~l} \bar{\Phi}_l, \\
\text{or equivalently} \quad 
D^{ij}_\A D^{mn}_\B \bar{\Phi}_k &= -i\del_{\A\B} \left( \e^{ijmn}\Phi^k 
                              - \e^{ljmn} \bar{\Phi}_l \D^i_k - \e^{ilmn} \bar{\Phi}_l \D^j_k
                              + \e^{ijln} \bar{\Phi}_l \D^m_k + \e^{ijml} \bar{\Phi}_l \D^n_k \right) \\
\end{split}
\ee

Using the above relation a straightforward computation shows that the following $R$-symmetry singlet integer 
spin currents are conserved
\be\label{N6curr}
J^{(s)}= \sum^s_{r=0} (-1)^r {2s \choose 2r} ~\del^r\bar\Phi_p ~\del^{s-r}\Phi^p 
           - \frac{1}{24}\sum^{s-1}_{r=0} (-1)^{r+1} {2s \choose 2r+1} ~\e_{ijkl}~\del^r D^{ij} \bar\Phi_p 
           ~\del^{s-r-1}D^{kl} \Phi^p. \\
\ee
where $\partial = i\lambda^\alpha\gamma^\mu_{\alpha\beta}\lambda^\beta\partial_\mu$, 
$D = \lambda^\alpha D_\alpha$ and $s = 0,1,2\ldots$. The stress-energy tensor of this 
theory lies, as in the ${\cal N}=4$ theory, in the R-symmetry singlet spin zero 
multiplet $(1,0,\{0,0,0\})$.

\section{Weakly broken conservation}\label{wbc}
The free superconformal theories discussed above have an exact higher spin symmetry 
algebra generated by the charges corresponding to the infinite number of conserved 
currents that these theories possess. These free theories can be deformed into interacting 
theories by turning on $U(N)(SU(N))$ Chern-Simons(CS) gauge interactions, 
in a supersymmetric fashion and preserving the conformal invariance of free CFTs, under 
which the matter fields transform in fundamental representations. The CS gauge interactions 
do not introduce any new local degrees of freedom so the spectrum of local operators in 
the theory remains unchanged. Turning on the interactions breaks the higher spin symmetry 
of the free theory but in a controlled way which we discuss below. These interacting 
CS vector models are interesting in there own right as non trivial interacting quantum 
field theories. Exploring the phase structure of these theories at finite temperature 
and chemical potential, provides a platform for studying a lot of interesting physics, 
at least in the large $N$ limit, using the techniques developed in \cite{Giombi:2011kc}. 

From a more string 
theoretic point of view, a very interesting example of this class of theories is 
the $U(N)\times U(M)$ ABJ theory in the vector model limit $\frac{M}{N}\rightarrow 0$. 
ABJ theory in this vector model limit has recently been argued to be holographically 
dual a non-abelian supersymmetric generalization of the non-minimal Vasiliev theory in 
$AdS_4$ \cite{Chang:2012kt}. The ABJ theory thus connects, as its holographic duals, Vasiliev 
theory at one end to a string theory at another end. Increasing $\frac{M}{N}$ from 0 
corresponds to increasing the coupling of $U(M)$ gauge interactions in the bulk Vasiliev 
theory. Thus, understanding the ABJ theory away from the vector model limit in an 
expansions in $\frac{M}{N}$ would be a first step towards understanding of how 
string theory emerges from `quantum' Vasiliev theory.\footnote{See \cite{Banerjee:2013nca} 
for a very recent attempt in this direction.}

In \cite{Maldacena:2011jn, Maldacena:2012sf} theories with exact conformal symmetry 
but weakly broken higher spin symmetry were studied. It was first observed in 
\cite{Giombi:2011kc}, and later used with great efficiency in \cite{Maldacena:2012sf}, 
that the anomalous ``conservation'' equations are of the schematic form 
\begin{equation}
\partial\cdot J_{(s)}=\frac{a}{N}J_{(s_{1})}J_{(s_{2})}+\frac{b}{N^{2}}J_{(s'_{1})}J_{(s'_{2})}J_{(s'_{3})}
\end{equation}
plus derivatives sprinkled appropriately. 
The structure of this equation is constrained on symmetry grounds -
the twist ($\Delta_{i}-s_{i}$) of the L.H.S. is $3$. If each $J_{s}$
has conformal dimension $\Delta=s+1+O(1/N)$, and thus twist $\tau=1+O(1/N)$,
the two terms on the R.H.S. are the only ones possible by twist matching.
Thus we can have only double or triple trace deformations in the case
of weakly broken conservation and terms with four or higher number
of currents are not possible.

In the superconformal case that we are dealing with, since $D$ has
dimension $1/2$ , $D\cdot J_{(s)}$ is a twist 2 operator. Thus in this
case the triple trace deformation is forbidden and the only possible
structure is more constrained:

\begin{equation}
D\cdot J_{(s)}=\frac{a}{N}J_{(s_{1})}J_{(s_{2})}
\end{equation}

In view of this, it is feasible that in large-N supersymmetric Chern-Simons 
theories the structure of correlation functions is much more constrained 
(compared to the non-supersymmetric case).

\section{Two-point functions}\label{2pf}
The two-point function of two spin-$s$ operators in a 3d SCFT has a form completely determined (upto overall multiplicative constants) by superconformal invariance. Since, as we saw in section 2, $X_{12\pm}$ is the only superconformally  covariant structure built out of two points in superspace, the only possible expression for the two point function which also has the right dimension and homogeneity in $\lambda$ is:
\begin{equation}\label{2ptfn}
\langle J_s(1) J_s(2)\rangle \propto \frac{P_3^{2s}}{\tilde{X}_{12}^2}
\end{equation}
where $P_3$ is the superconformal invariant defined on two points, given in Table \ref{invartab}.
The overall constant can be determined in free field theory, see below.

As an illustrative example, we  consider the two-point function of two spin half supercurrents. On the basis of symmetry and dimension matching we can have the following
possible structure for the 2-point function:

\begin{equation}
\langle J_{1/2}(x_{1},\theta_{1},\lambda_1)J_{1/2}(x_{2},\theta_{2},\lambda_2)\rangle=b\frac{\lambda_{1}\lambda_{2}}{\tilde{X}_{12}^{\Delta_{1}+\Delta_{2}}}\frac{\theta_{12}^{2}}{\tilde{X}_{12}}+\frac{\lambda_{1}\tilde{X}_{12}\lambda_{2}}{\tilde{X}_{12}^{\Delta_{1}+\Delta_{2}+1}}(c+d\frac{\theta_{12}^{2}}{\tilde{X}_{12}})
\end{equation}
where $\tilde{X}_{12}\equiv\sqrt{(\tilde{X}_{12})_{\alpha}^{\,\,\beta}(\tilde{X}_{12})_{\beta}^{\,\,\alpha}}$ \footnote{note that throughout $\tilde{X}_{12}$ denotes this scalar object. The matrix will always be denoted  with the indices: $(\tilde{X}_{12})_{\alpha}^{\,\beta}$ }.
The shortening condition on the above 2-point function gives

\begin{equation}
d=0\,\,\,\,\,\,\,\,\,b=\frac{ic}{4}(\Delta_{1}+\Delta_{2}-2)
\end{equation}
For $J_{1/2}$ a superconformal primary $\Delta_{1}=\Delta_{2}=3/2$ so $b=ic/4$ and the two point function (upto some undetermined overall normalization) is given by

\begin{equation}
\langle J_{1/2}(x_{1},\theta_{1},\lambda_1)J_{1/2}(x_{2},\theta_{2},\lambda_2)\rangle\propto\frac{\lambda_{1}\tilde{X}_{12}\lambda_{2}}{\tilde{X}_{12}^{4}}+\frac{i}{4}\frac{\lambda_{1}\lambda_{2}\theta_{12}^{2}}{\tilde{X}_{12}^{4}}
\end{equation}
A natural  generalization, that reduces correctly to the above equation for $s=1/2$, is

\begin{equation}
\langle J_{s}(1)J_{s}(2)\rangle\propto\frac{(\lambda_{1}\tilde{X}_{12}\lambda_{2})^{2s-1}}{\tilde{X}_{12}^{4s+2}}(\lambda_{1}\tilde{X}_{12}\lambda_{2}+\frac {is}{2}\lambda_{1}\lambda_{2}\theta_{12}^{2})\,\,\,
\end{equation}
with $\langle J_{0}J_{0}\rangle=1/\tilde{X}_{12}^{2}$
(since the superconformal shortening condition is different for
spin zero). Note that the above can be written as

\begin{equation}
\langle J_{s}(1)J_{s}(2)\rangle\propto\frac{(\lambda_{1}\tilde{X}_{12}\lambda_{2}+\frac {i}{4}\lambda_{1}\lambda_{2}\theta_{12}^{2})^{2s}}{\tilde{X}_{12}^{4s+2}}
\end{equation}
which is the same as \eqref{2ptfn}. The shortening condition on this is satisfied, as may be explicitly checked.

As a check, we also work out,  by elementary field theory methods, the two point function of the spin $\half$ current constructed out of the free ${\cal N}=1$
superfield which is defined as\footnote{We insert a factor of $i$ in this definition for convenience, which differs
from the definition given in \eqref{freef}.}
\begin{equation}
\begin{split}
\Phi &= \phi + i\theta\psi \\
{\bar\Phi} &= {\bar\phi} + i\theta\psi^*
\end{split}
\end{equation}

We find that the 2-point function computed explicitly in the free theory is in agreement with our result \eqref{2ptfn} above.
The spin half supercurrent is
\begin{equation}
J_{\alpha}={\bar\Phi}D_{\alpha}\Phi - (D_{\alpha}{\bar\Phi})\Phi
\end{equation}
Using the equation of motion for $\Phi$ this obeys the shortening
condition $D^{\alpha}J_{\alpha}=0$. The two point function of two
such currents can be obtained after doing Wick contractions to write 4-point functions
in terms of 2-point functions. We use the free field propagator $\langle{\bar\Phi}\Phi\rangle = \frac{1}{{\tilde X}_{12}}$,
and also that,
\begin{equation}
D_{1\alpha}D_{2\beta}\frac{1}{\tilde{X}_{12}}=\frac{-i(\tilde{X}_{12})_{\alpha\beta}}{(\tilde{X}_{12})^{3}}\,\,\,\,\,\,\,\,\,\,,\,\,\,\,\, D_{1\alpha}\frac{1}{\tilde{X}_{12}}D_{2\beta}\frac{1}{\tilde{X}_{12}}=\frac{\epsilon_{\alpha\beta}\theta_{12}^{2}}{4(\tilde{X}_{12})^{4}}
\end{equation}
This gives (upto multiplicative factors which we neglect)
\begin{equation}
\begin{split}
 \langle J_\alpha(1) J_\beta(2)\rangle &\propto \frac{((-{\tilde X}_{12})_{\alpha\beta} + \frac{i}4\theta_{12}^2\epsilon_{\alpha\beta})}{{\tilde X}_{12}^4}.
\end{split} 
\end{equation}
Contracting with $\lambda_1^\alpha$ and $\lambda_2^\beta$ we find, in free field theory,
\begin{equation}
\begin{split}
 \langle J_\half(1)J_\half(2)\rangle &= -i\frac{P_3}{\tilde{X}_{12}^2}
\end{split} 
\end{equation}
which, indeed, is what was expected. One can determine the constants appearing in front of the
two point function in free field theory and we divide by it (so that the final result is 
normalized to one) which gives the following result for
general spin $s$
\begin{equation}
 \langle J_s(1)J_s(2)\rangle = c(s)\frac{P_3^{2s}}{\tilde{X}_{12}^2}
\end{equation}
where $c(s) = \left(\frac{i}{2}\right)^{2s} \frac{\sqrt{\pi}}{s!\Gamma(s + \half)}$
for all $s\geq0.$

\section{Three-point functions}\label{3pf}
In this section we undertake the task of determining all the possible
structures that can occur in the three-point functions of higher spin
operators $\langle J_{s_{1}}J_{s_{2}}J_{s_{3}}\rangle$. For
the nonsupersymmetric case this was done in \cite{Giombi:2011rz}.  We will use superconformal invariance
to ascertain what structures can occur
in three-point functions. 

The structure of correlation functions in SCFTs has been earlier
studied by J-H Park \cite{Park:1999cw, Park:1997bq, Park:1998nra, Park:1999pd} and H. Osborn \cite{Osborn:1998qu}.\footnote{Kuzenko \cite{Kuzenko:1999pi}
has also studied 3-pt functions of the supercurrent and flavour currents of ${\cal N}=2$ 4d SCFTs.}.
The structure of covariant objects which are used as building blocks for
the construction of invariants in the present work was entirely laid out in the above references.
However, our goal in the present work is to make use of these structures to study theories
which have conserved currents of higher spin. For this purpose, it is convenient to adopt
the polarization spinor formalism of \cite{Giombi:2011rz}. After writing down the structures that
can appear for a given three-point function, we use on-shell conservation laws of the currents
to constrain the coefficients appearing in front of the structures.
% We build on their results on the supercovariant structures in superspace (reviewed in section 2) and use them
% together with the polarization spinor formalism of \cite{Giombi:2011rz} to carry out our analysis.

We find that there exist new structures
for both the parity even and odd part of $\langle J_{s_{1}}J_{s_{2}}J_{s_{3}}\rangle$
which were not present in the nonsupersymmetric case. The parity-odd superconformal invariants are of special interest
as they arise in interacting 3d SCFTs. We will here restrict ourselves to the case of $\mathcal{N}=1$ SCFTs (no $R$-symmetry).
The results are summarized in the table given below:
\begin{table}[h]
\centering
\begin{tabular}{| c | c | c |}
\hline
            & Parity even                                                   & Parity odd \\
\hline
 Bosonic    & $P_{1}=\lambda_{2}X_{23-}^{-1}\lambda_{3}$                    &            \\
            &                                                               &  $S_{1}=\frac{\lambda_{3}X_{31+}X_{12+}\lambda_{2}}{\tilde{X}_{12}\tilde{X}_{23}\tilde{X}_{31}}$          \\
            & $Q_{1}=\lambda_{1}X_{12-}^{-1}X_{23+}X_{31-}^{-1}\lambda_{1}$ &  and cyclic            \\
            & and cyclic &\\
\hline 
            &                                                               & \\  
 Fermionic  & $R_{1}=\lambda_{1}\Theta_{1}$ and cyclic                      &   $T=\tilde{X}_{31}\frac{\Theta_{1}X_{12+}X_{23+}\Theta_{3}}{\tilde{X}_{12}\tilde{X}_{23}}$\\
            & &\\  
\hline
\end{tabular}
\caption{Invariant structures in ${\cal N}=1$ superspace.}\label{invartab}
\end{table}

\subsection{Superconformal invariants for three-point functions of $\mathcal{N}=1$
higher spin operators}\label{N13pfinv}

We need to determine all the superconformal invariants that can be
constructed out of the co-ordinates of (augmented) superspace : $x_{i},\,\theta_{i}$
and the (bosonic) polarization spinors $\lambda_{i}$ ($i=1,2,3$). Using the covariant objects of section 2, which transformed homogeneously under superinversions, we can
begin to write down the superconformal invariants constructed out
of ($x_{i},\,\theta_{i},\,\lambda_{i}$).

We have \begin{equation}
\lambda_{i}X_{ij-}^{-1}\lambda_{j}\rightarrow-(\lambda_{i}X_{i-}^{-1})(-X_{i-}X_{ij-}^{-1}X_{j-})(X_{j+}^{-1}\lambda_{j})=\lambda_{i}X_{ij-}^{-1}\lambda_{j}\end{equation}
 Thus we have the three superconformal invariants\begin{equation}\label{pdefs}
P_{1}=\lambda_{2}X_{23-}^{-1}\lambda_{3}\,,\,\,\,\,\, P_{2}=\lambda_{3}X_{31-}^{-1}\lambda_{1}\,,\,\,\,\,\, P_{3}=\lambda_{1}X_{12-}^{-1}\lambda_{2}\end{equation}
 Also, under superinversion,\begin{equation}
\mathfrak{X}_{1+}=X_{12-}^{-1}X_{23+}X_{31-}^{-1}\rightarrow-X_{1-}\mathfrak{X}_{1+}X_{1+}\end{equation}
and similarly for $\mathfrak{X}_{2+},\mathfrak{X}_{3+}$, so we also
have the following as superconformal invariants:\begin{equation}
Q_{1}=\lambda_{1}\mathfrak{X}_{1+}\lambda_{1}\,,\,\,\,\,\, Q_{2}=\lambda_{2}\mathfrak{X}_{2+}\lambda_{2}\,,\,\,\,\,\, Q_{3}=\lambda_{3}\mathfrak{X}_{3+}\lambda_{3}\end{equation}
 Furthermore, \begin{equation}
\lambda_{3}X_{31+}X_{12+}\lambda_{2}\rightarrow-\frac{1}{x_{1}^{2}x_{2}^{2}x_{3}^{2}}\lambda_{3}X_{31+}X_{12+}\lambda_{2}\,,\,\,\,\,\,\tilde{X}_{ij}^{2}\rightarrow\frac{\tilde{X}_{ij}^{2}}{x_{i}^{2}x_{j}^{2}}\end{equation}
 so there are the additional (parity odd) superconformal invariants\begin{equation}
S_{1}=\frac{\lambda_{3}X_{31+}X_{12+}\lambda_{2}}{\tilde{X}_{12}\tilde{X}_{23}\tilde{X}_{31}}\,,\,\,\,\,\, S_{2}=\frac{\lambda_{1}X_{12+}X_{23+}\lambda_{3}}{\tilde{X}_{12}\tilde{X}_{23}\tilde{X}_{31}}\,,\,\,\,\,\, S_{3}=\frac{\lambda_{2}X_{23+}X_{31+}\lambda_{1}}{\tilde{X}_{12}\tilde{X}_{23}\tilde{X}_{31}}\end{equation}
which transform to minus themselves under inversion. Together these
constitute the supersymmetric generalizations of the conformally invariant
$P,\, Q,\, S$ structures discussed in \cite{Giombi:2011rz} %
\footnote{Note that the $S_{k}$ in \cite{Giombi:2011rz} has an extra factor of $i P_{k}$
compared to ours. %
}

 Using the covariant $\Theta$ structures of section 2 it follows that we have the additional (parity even) fermionic invariants\begin{equation}
R_{1}=\lambda_{1}\Theta_{1}\,,\,\,\,\,\, R_{2}=\lambda_{2}\Theta_{2}\,,\,\,\,\,\, R_{3}=\lambda_{3}\Theta_{3}\end{equation}
 It may be checked that \begin{equation}
R_{1}^{2}=R_{2}^{2}=R_{3}^{2}=R_{1}R_{2}R_{3}=0\end{equation}

\subsubsection{Construction of the parity odd fermionic invariant $T$}\label{3pfpo}

We can construct more superconformally covariant structures from the
building blocks $(X_{jk+},\,\mathfrak{X}_{i+},\,\Theta_{i},\,\lambda_{i})$
- these are the fermionic analogues of $P,\, S,\, Q$. We define them
below and also give there transformation under superinversion.

a) Fermionic analogues of $P_{i}$: Define \begin{equation}
\pi_{ij}=\lambda_{i}X_{ij+}\Theta_{j}\end{equation}
 Then under superinversion\begin{equation}
\pi_{ij}\rightarrow-\lambda_{i}X_{i-}^{-1}X_{i+}^{-1}X_{ij+}X_{j-}^{-1}X_{j-}\Theta_{j}=-\frac{1}{x_{i}^{2}}\pi_{ij}\end{equation}
 Similarly,

\begin{equation}
\Pi_{ij}=\Theta_{i}X_{ij+}\Theta_{j}\,,\,\,\,\,\,\,\Pi_{ij}\rightarrow\Pi_{ij}\end{equation}
 It turns out, however, that \begin{equation}
\Pi_{ij}=0\end{equation}

b) Fermionic analogues of $S_{i}$:\begin{equation}
\sigma_{13}=\frac{\lambda_{1}X_{12+}X_{23+}\Theta_{3}}{\tilde{X}_{12}\tilde{X}_{23}\tilde{X}_{31}},\,\,\,\,\,\,\,\sigma_{13}\rightarrow x_{3}^{2}\sigma_{13}\end{equation}
\begin{equation}
\Sigma_{13}=\frac{\Theta_{1}X_{12+}X_{23+}\Theta_{3}}{\tilde{X}_{12}\tilde{X}_{23}\tilde{X}_{31}}\,,\,\,\,\,\,\,\,\Sigma_{13}\rightarrow-x_{1}^{2}x_{3}^{2}\Sigma_{13}\end{equation}
 $\sigma_{32},\sigma_{21},\Sigma_{32},\Sigma_{21}$ are similarly
defined through cyclic permutation of the indices. It follows that\begin{equation}
\tilde{X}_{ij}^{2}\Sigma_{ij}\rightarrow-\tilde{X}_{ij}^{2}\Sigma_{ij}\end{equation}

c) Fermionic analogues of $Q_{i}$:

\begin{equation}
\omega_{i}=\lambda_{i}\mathfrak{X}_{i+}\Theta_{i}\,,\,\,\,\,\,\,\,\omega_{i}\rightarrow-x_{i}^{2}\omega_{i}\end{equation}
\begin{equation}
\Omega_{i}=\Theta_{i}\mathfrak{X}_{i+}\Theta_{i}\,,\,\,\,\,\,\,\,\Omega_{i}\rightarrow x_{i}^{4}\Omega_{i}\end{equation}
 However, $\Omega_{i}$ is identically zero\begin{equation}
\Omega_{i}=0\end{equation}

The invariants constructed out of the product of two parity odd (or
two parity even) covariant structures would be parity even, and since
we have already listed all the parity even invariants, would be expressible
in terms of $P_{i},\, Q_{i},\, R_{i}.$ Thus, we find the following
relations for the above covariant structures\begin{equation}
\pi_{ij}^{2}=\sigma_{ij}^{2}=\omega_{i}^{2}=0\end{equation}
\begin{equation}
\pi_{ij}\omega_{i}=0\end{equation}
\begin{equation}
\frac{1}{\tilde{X}_{12}^{2}}\pi_{12}\pi_{23\,}=-R_{1}R_{2}\,,\,\,\,\,\,\frac{1}{\tilde{X}_{23}^{2}}\pi_{23}\pi_{31\,}=-R_{2}R_{3}\,,\,\,\,\,\,\frac{1}{\tilde{X}_{31}^{2}}\pi_{31}\pi_{12\,}=-R_{3}R_{1}\end{equation}

\begin{equation}
\frac{1}{\tilde{X}_{ij}^{2}}\pi_{ij}\pi_{ji\,}=R_{i}R_{j}=\tilde{X}_{ij}^{2}\sigma_{ij}\sigma_{ji}\end{equation}
\begin{equation}
\tilde{X}_{12}^{2}\sigma_{21}\sigma_{32}=R_{2}R_{3}\,,\,\,\,\,\,\tilde{X}_{23}^{2}\sigma_{32}\sigma_{13}=R_{3}R_{1}\,,\,\,\,\,\,\tilde{X}_{31}^{2}\sigma_{13}\sigma_{21}=R_{1}R_{2}\end{equation}
\begin{equation}
\tilde{X}_{ij}^{2}\,\omega_{i}\omega_{j}=-R_{i}R_{j}\end{equation}

From the above covariant structures it is possible to build additional
parity odd fermionic invariants by taking products of a parity even
and a parity odd covariant structure.%
\footnote{Note that structures like $x_{i}\omega_{i},\,\pi_{ij}/x_{i}$ would
be parity odd invariants under inversion. However, these are not Poincare
invariant (since correlation functions should depend only on differences
($x_{ij}$) of the coordinates). We could also construct structures
like $U=\tilde{X}_{12}\tilde{X}_{23}\tilde{X}_{31}\omega_{1}\omega_{2}\omega_{3}$
which would be an odd invariant ($U\rightarrow-U$) . However, it
is identically zero because the product of three different $\Theta$'s
vanishes. %
}Thus, we have\begin{equation}
T_{ij}=\pi_{ij}\sigma_{ji}\end{equation}
and under superinversion\begin{equation}
T_{ij}\rightarrow-T_{ij}\end{equation}
 Note that $\pi_{ij}\ne\pi_{ji}$ so $\{\pi_{12},\,\pi_{23},\,\pi_{31}\}$
is a different set of parity odd covariant structures than $\{\pi_{21},\,\pi_{32},\,\pi_{13}\}$
(the same is true for the even structures $\sigma_{ij}$). However,
because the following relation is true\begin{equation}
T_{ij}=-T_{ji}\end{equation}
it follows that we have only three odd invariant structures:\begin{equation}
T_{1}\equiv T_{23}=\pi_{23}\sigma_{32}\,,\,\,\,\,\, T_{2}\equiv T_{31}=\pi_{31}\sigma_{13}\,,\,\,\,\,\, T_{3}\equiv T_{12}=\pi_{12}\sigma_{21}\end{equation}

We may also define\begin{equation}
T'_{23}=\pi_{12}\sigma_{31}\,,\,\,\,\,\, T'_{31}=\pi_{23}\sigma_{12}\,,\,\,\,\,\, T'_{12}=\pi_{31}\sigma_{23}\,\,\,\,\,\,\, T'_{ij}\rightarrow-T'_{ij}\end{equation}
 with $T'_{32}=\pi_{13}\sigma_{21}\,,\,\,\,\,\, T'_{13}=\pi_{21}\sigma_{32}\,,\,\,\,\,\, T'_{21}=\pi_{32}\sigma_{13}$
again being related to the above by \begin{equation}
P_{3}T'_{21}=-P_{2}T'_{31}\,,\,\,\,\,\, P_{1}T'_{32}=-P_{3}T'_{12}\,,\,\,\,\,\, P_{2}T'_{13}=-P_{1}T'_{23}\end{equation}

Also\begin{equation}
\bar{T}_{ij}=\tilde{X}_{ij}^{2}\sigma_{ji}\omega_{j}\,,\,\,\,\,\,\,\,\bar{T}_{ij}\rightarrow-\bar{T}_{ij}\end{equation}
 Again, we have the relation\begin{equation}
\bar{T}_{ij}Q_{i}=\bar{T}_{ji}Q_{j}\end{equation}
 thus we have only three $\bar{T}_{ij}$'s. 

Likewise, we have \begin{equation}
\hat{T}_{12}=\tilde{X}_{12}^{2}\sigma_{31}\omega_{2}\,,\,\hat{T}_{23}=\tilde{X}_{23}^{2}\sigma_{12}\omega_{3}\,\hat{T}_{31}=\tilde{X}_{31}^{2}\sigma_{23}\omega_{1}\,\,\,\,\,\,\hat{T}_{ij}\rightarrow-\hat{T}_{ij}\end{equation}
 with $\hat{T}_{21}\,,\hat{T}_{32}\,,\hat{T}_{13}$ being related
to the above by\begin{equation}
P_{j}\hat{T}_{ij}=P_{i}\hat{T}_{ji}\end{equation}

We also have the following relations involving $\Sigma_{ij}$\begin{equation}
\Sigma_{ij}=\Sigma_{ji}\,,\,\,\,\,\,\,\,\tilde{X}_{12}^{2}\Sigma_{12}=\tilde{X}_{23}^{2}\Sigma_{32}=\tilde{X}_{31}^{2}\Sigma_{31}\end{equation}
Therefore, here we get just one parity odd invariant\begin{equation}
T\equiv\tilde{X}_{ij}^{2}\Sigma_{ij}\label{eq:Tdef}\end{equation}

It turns out that $T'_{ij},\,\bar{T}_{ij},\,\hat{T}_{ij}\,,\tilde{X}_{ij}^{2}\Sigma_{ij}$
can be expressed in terms of $T_{i}$ by means of the following relations\[
P_{1}T'_{31}=P_{3}T_{1}\,,\,\,\,\,\, P_{2}T'_{12}=P_{1}T_{2}\,,\,\,\,\,\, P_{3}T'_{23}=P_{2}T_{3}\]
\begin{equation}
P_{3}\bar{T}_{12}=-Q_{2}T_{3}\,,\,\,\,\,\, P_{1}\bar{T}_{23}=-Q_{3}T_{1}\,,\,\,\,\,\, P_{2}\bar{T}_{31}=-Q_{1}T_{2}\end{equation}
\[
\frac{1}{2}P_{2}\tilde{X}_{13}^{2}\Sigma_{13}=T_{2}\,,\,\,\,\,\,\frac{1}{2}P_{3}\tilde{X}_{21}^{2}\Sigma_{21}=T_{3}\,,\,\,\,\,\,\frac{1}{2}P_{1}\tilde{X}_{32}^{2}\Sigma_{32}=T_{1}\]
\[
P_{1}\hat{T}_{23}=-P_{2}T_{1}\,,\,\,\,\,\, P_{2}\hat{T}_{31}=-P_{3}T_{2}\,,\,\,\,\,\, P_{3}\hat{T}_{12}=-P_{1}T_{3}\]
Making use of the above equation and eq.(\ref{eq:Tdef}) we can express
all parity odd fermionic structures in terms of $T$

\begin{equation}
T_{2}=\frac{1}{2}P_{2}T\,,\,\,\,\,\, T_{3}=\frac{1}{2}P_{3}T\,,\,\,\,\,\, T_{1}=\frac{1}{2}P_{1}T\end{equation}

\begin{equation}
\bar{T}_{12}=-\frac{1}{2}Q_{2}T\,,\,\,\,\,\,\bar{T}_{23}=-\frac{1}{2}Q_{3}T\,,\,\,\,\,\,\bar{T}_{31}=-\frac{1}{2}Q_{1}T\end{equation}
\begin{equation}
T'_{31}=\frac{1}{2}P_{3}T\,,\,\,\,\,\, T'_{12}=\frac{1}{2}P_{1}T\,,\,\,\,\,\, T'_{23}=\frac{1}{2}P_{2}T\end{equation}
\begin{equation}
\hat{T}_{12}=-\frac{1}{2}P_{1}T\,,\,\,\,\,\,\hat{T}_{23}=-\frac{1}{2}P_{2}T\,,\,\,\,\,\,\hat{T}_{31}=-\frac{1}{2}P_{3}T\end{equation}

To summarize, from our fermionic covariant structures we could construct
five parity odd invariants $T_{i},\, T'_{ij},\,\bar{T}_{ij},\,\hat{T}_{ij}\,,T$.
However, only $T$ suffices as the other four are related to it through
the above simple relations. 

{\it Summary of this section}: \,\,\,We have thus obtained the superconformal
invariants\\
 $P_{i},\, Q_{i},\, R_{i},\, S_{i},\, T$ (listed in tabular form at the beginning of this section) out of which the
invariant structures for particular 3-point functions can be constructed
as monomials in these variables. Before we do this, however, we need
to determine all the relations between these variables using which
we can get a linearly independent basis of monomial structures for
3-point functions.

\subsection{Relations between the invariant structures}\label{relns}

Since the $\mathcal{N}=1$ superconformal group in 3 dimensions has
14 generators (10 bosonic, 4 fermionic), out of ($x_{i},\,\theta_{i},\,\lambda_{i}$)
($i=1,2,3$) we can construct $7\times3-14=7$ superconformal invariants.
Thus among the nine parity even structures $(P_{i}\,,Q_{i}\,,R_{i})$
we must have two relations. One of them is the supersymmetrized version
of the non-linear relation (2.14) in \cite{Giombi:2011rz} \begin{equation}
P_{1}^{2}Q_{1}+P_{2}^{2}Q_{2}+P_{3}^{2}Q_{3}-2P_{1}P_{2}P_{3}-Q_{1}Q_{2}Q_{3}-\frac{i}{2}(R_{1}R_{2}P_{3}Q_{3}+R_{2}R_{3}P_{1}Q_{1}+R_{3}R_{1}P_{2}Q_{2})=0\label{eq:nl1}\end{equation}
This cuts down the number of independent invariants by one. We also
have the following triplet of relations which vanishes identically
when the Grassmann variables are set to zero (fermionic relations)
and reduces the number of invariants to seven :\[
P_{2}R_{1}R_{2}+Q_{1}R_{2}R_{3}+P_{3}R_{3}R_{1}=0\]
\begin{equation}
P_{3}R_{2}R_{3}+Q_{2}R_{3}R_{1}+P_{1}R_{1}R_{2}=0\label{eq:nl2}\end{equation}
\[
P_{1}R_{3}R_{1}+Q_{3}R_{1}R_{2}+P_{2}R_{2}R_{3}=0\]
There are further non-linear relations involving the $S$'s. Since
the squares or products of $S$'s are parity even, we expect them
to be determined in terms of the parity even structures. Indeed, we
find\begin{equation}
S_{1}^{2}=P_{1}^{2}-Q_{2}Q_{3}-iP_{1}R_{2}R_{3}\,,\,\,\,\,\, S_{2}^{2}=P_{2}^{2}-Q_{3}Q_{1}-iP_{2}R_{3}R_{1}\,,\,\,\,\,\, S_{3}^{2}=P_{3}^{2}-Q_{1}Q_{2}-iP_{3}R_{1}R_{2}\label{eq:nl3}\end{equation}
\[
S_{1}S_{2}=P_{3}Q_{3}-P_{1}P_{2}\,,\,\,\,\,\, S_{2}S_{3}=P_{1}Q_{1}-P_{2}P_{3}\,,\,\,\,\,\, S_{3}S_{1}=P_{2}Q_{2}-P_{3}P_{1}\]
They imply that the most general odd structures that can occur in
any three point function are linear in $S_{i}.$ It turns out there
exist further {\it linear} relations between the parity odd structures.
We find the following basic linear relationships between the various
parity odd invariant structures:

At $O(\lambda_{1}\lambda_{2}\lambda_{3})$: \begin{equation}
R_{1}S_{1}+R_{2}S_{2}+R_{3}S_{3}=0\label{eq:nl4}\end{equation}

At $O(\lambda_{1}^{2}\lambda_{2}\lambda_{3},\,\lambda_{1}\lambda_{2}^{2}\lambda_{3},\,\lambda_{1}\lambda_{2}\lambda_{3}^{2})$:
\[
Q_{1}S_{1}+P_{2}S_{3}+P_{3}S_{2}-\frac{i}{2}P_{2}P_{3}T=0\]
\begin{equation}
Q_{2}S_{2}+P_{3}S_{1}+P_{1}S_{3}-\frac{i}{2}P_{1}P_{3}T=0\label{eq:nl5}\end{equation}
\[
Q_{3}S_{3}+P_{1}S_{2}+P_{2}S_{1}-\frac{i}{2}P_{1}P_{2}T=0\]

and \[
S_{2}R_{1}R_{2}+S_{3}R_{3}R_{1}+T(Q_{1}P_{1}-P_{2}P_{3})=0\]
\begin{equation}
S_{3}R_{2}R_{3}+S_{1}R_{1}R_{2}+T(Q_{2}P_{2}-P_{3}P_{1})=0\label{eq:nl6}\end{equation}
\[
S_{1}R_{3}R_{1}+S_{2}R_{2}R_{3}+T(Q_{3}P_{3}-P_{1}P_{2})=0\]
 From eq. (\ref{eq:nl4}) follows:\[
S_{2}R_{1}R_{2}-S_{3}R_{3}R_{1}=0\]
\begin{equation}
S_{3}R_{2}R_{3}-S_{1}R_{1}R_{2}=0\label{eq:nl7}\end{equation}
\[
S_{1}R_{3}R_{1}-S_{2}R_{2}R_{3}=0\]
 From these follow other linear relations at higher orders in $\lambda_{1}\,,\lambda_{2}\,,\lambda_{3}$:
\[
Q_{1}P_{1}S_{1}+Q_{2}P_{2}S_{2}-Q_{3}P_{3}S_{3}+2P_{1}P_{2}S_{3}-\frac{i}{2}TP_{1}P_{2}P_{3}=0\]
\begin{equation}
Q_{2}P_{2}S_{2}+Q_{3}P_{3}S_{3}-Q_{1}P_{1}S_{1}+2P_{2}P_{3}S_{1}-\frac{i}{2}TP_{1}P_{2}P_{3}=0\end{equation}
\[
Q_{3}P_{3}S_{3}+Q_{1}P_{1}S_{1}-Q_{2}P_{2}S_{2}+2P_{3}P_{1}S_{2}-\frac{i}{2}TP_{1}P_{2}P_{3}=0\]
Adding the above equations gives\begin{equation}
 Q_{1}P_{1}S_{1}+Q_{2}P_{2}S_{2}+Q_{3}P_{3}S_{3}-\frac{3i}{2}TP_{1}P_{2}P_{3}+2(P_{1}P_{2}S_{3}+P_{2}P_{3}S_{1}+P_{3}P_{1}S_{2})=0\end{equation}

 Also, we get \begin{equation}
R_{1}R_{2}(S_{1}P_{2}+\frac{1}{2}Q_{3}S_{3})+R_{2}R_{3}(S_{2}P_{3}+\frac{1}{2}Q_{1}S_{1})+R_{3}R_{1}(S_{3}P_{1}+\frac{1}{2}Q_{2}S_{2})=0\end{equation}

\[
(P_{1}^{2}Q_{1}-P_{2}^{2}Q_{2})P_{3}S_{3}+(P_{3}^{2}-Q_{1}Q_{2}-iP_{3}R_{1}R_{2})(Q_{1}P_{1}S_{1}-Q_{2}P_{2}S_{2})=0\]
\begin{equation}
(P_{2}^{2}Q_{2}-P_{3}^{2}Q_{3})P_{1}S_{1}+(P_{1}^{2}-Q_{2}Q_{3}-iP_{1}R_{2}R_{3})(Q_{2}P_{2}S_{2}-Q_{3}P_{3}S_{3})=0\label{eq:nl8}\end{equation}
\[
(P_{3}^{2}Q_{3}-P_{1}^{2}Q_{1})P_{2}S_{2}+(P_{2}^{2}-Q_{3}Q_{1}-iP_{2}R_{3}R_{1})(Q_{3}P_{3}S_{3}-Q_{1}P_{1}S_{1})=0\]
and so on. All these relations can be put to use in eliminating linearly
dependent structures in 3-point functions. The above relations between
the invariant structures extend the corresponding non-supersymmetric
ones in \cite{Giombi:2011rz}.

We also have the following relations\begin{equation}
T^{2}=0\,,\,\,\,\,\, FT=0\,,\,\,\,\,\, S_{i}T=-\epsilon_{ijk}R_{j}R_{k}\,\,\,\,\,\,\, sum\, over\, j,k\label{eq:nl9}\end{equation}
where $F$ stands for any of the fermionic covariant/invariant structures.
This implies that for any 3-point function it suffices to consider
parity odd structures linear in $T,\, S_{i}$. Thus $S_{i},\, T$
comprise all the parity odd invariants we need in writing down possible
odd structures in the 3-point functions of higher spin operators and
we need only terms linear in these invariants.

\subsection{Simple examples of three point functions}\label{exp3pf}

\subsubsection{Independent invariant structures for three point functions}\label{indep}
Below we write down the possible superconformal invariant structures
that can occur in specific three point functions $\langle J_{s_{1}}(1)J_{s_{2}}(2)J_{s_{3}}(3)\rangle$.
We consider the case of abelian currents so that, when some spins
are equal, the correlator is (anti-) symmetric under pairwise exchanges
of identical currents. We use only superconformal invariance to constrain
the correlators, so the results of this section apply even if the
higher spin symmetry is broken (that is, if $J_{s}$ is not conserved
for $s>2$). All that is required is that $J_{s}$ are higher spin
operators transforming suitably under superconformal transformations%
\footnote{We take $J_{\alpha_{1}\alpha_{2}.....\alpha_{s_{i}}}$ to be a primary
with arbitrary conformal dimension $\Delta_{i}$ so that $J_{s_{i}}\equiv\lambda^{\alpha_{1}}\lambda^{\alpha_{2}}...\lambda^{\alpha_{s_{i}}}J_{\alpha_{1}\alpha_{2}.....\alpha_{s_{i}}}$
has dimension $\Delta_{i}-s_{i}$. In general $J_{s_{i}}$ need not
be conserved. However, if the unitarity bound is attained - $\Delta_{i}=s_{i}+1$
for $s_{i}\ge\frac{1}{2}$ ; $\Delta_{i}=\frac{1}{2}$ for $s_{i}=0$-
then $J_{s_{i}},$ being a short primary, is necessarily conserved:
$D_{(i)\alpha}\frac{\partial}{\partial\lambda_{(i)\alpha}}J_{s_{i}}=0$%
}.

Under the pairwise exchange $2\leftrightarrow3$ we have\begin{equation}
A_{1}\rightarrow-A_{1}\,,\,\,\,\,\, A_{2}\rightarrow-A_{3}\,,\,\,\,\,\, A_{3}\rightarrow-A_{2}\,,\,\,\,\,\, T\rightarrow T\end{equation}
 where $A$ stands for any of $P,\, Q,\, R,\, S$.

$\langle J_{\frac{1}{2}}J_{\frac{1}{2}}J_{0}\rangle$: Here $J_{0}$
is a scalar operator with $\Delta=1$. It is clear that any term that
can occur is of order $\lambda_{1}\lambda_{2}$. Thus the possible
structures that can occur in this correlator are:\begin{equation}
P_{3}\,,\, R_{1}R_{2}\,,\, S_{3}\,,\, P_{3}T\end{equation}
 We also computed this correlator explicitly in the free field theory
(like the $\langle J_{\frac{1}{2}}J_{\frac{1}{2}}\rangle$ correlator
in the previous section) and the result is (with $\Delta_{1}=\Delta_{2}=\frac{3}{2}\,,\,\,\Delta_{3}=\frac{1}{2}$):
\begin{equation}
\frac{1}{\tilde{X}_{12}\tilde{X}_{23}\tilde{X}_{31}}(P_{3}-\frac{i}{2}R_{1}R_{2})\end{equation}
 The odd piece can not occur in the free field case. 

$\langle J_{\frac{1}{2}}J_{\frac{1}{2}}J_{\frac{1}{2}}\rangle$: Note
that this has to be antisymmetric under exchange of any two currents.
However the only two possible structures $\sum R_{i}P_{i}\,,\,\sum R_{i}S_{i}$
are symmetric under this exchange. Thus $\langle J_{\frac{1}{2}}J_{\frac{1}{2}}J_{\frac{1}{2}}\rangle$
vanishes.

$\langle J_{s}J_{0}J_{0}\rangle$ : For $s$ an even integer, the
correlator is

\begin{equation}
\langle J_{s}J_{0}J_{0}\rangle=\frac{1}{\tilde{X}_{12}\tilde{X}_{23}\tilde{X}_{31}}Q_{1}^{s}\end{equation}
 In this case no other structure can occur. For s odd or half-integral,
the correlator is zero.

$\langle J_{s}J_{\frac{1}{2}}J_{\frac{1}{2}}\rangle$: For $s$ an
even integer, the possible structures are

\begin{eqnarray*}
Q_{1}^{s}P_{1}\,,\, Q_{1}^{s-1}P_{2}P_{3}\,,\, R_{2}R_{3}Q_{1}^{s}\,,\\
Q_{1}^{s-1}(P_{2}S_{3}+P_{3}S_{2})\,,\, Q_{1}^{s}P_{1}T\,,\, Q_{1}^{s-1}P_{2}P_{3}T\,\end{eqnarray*}
The structure $R_{1}Q_{1}^{s-1}(R_{2}P_{2}-R_{3}P_{3})$ is also possible
but using eq.(\ref{eq:nl2}) equals $-R_{2}R_{3}Q_{1}^{s}$ and hence
can be eliminated while writing down independent superconformal invariant
structures. Similarly, the structure $Q_{1}^{s}S_{1}$ can be written
in terms of others listed above by using eq. (\ref{eq:nl5}) and $ R_{1}Q_{1}^{s-1}(R_{2}S_{2}-R_{3}S_{3})$ in terms of the last two structures above by using  eq. (\ref{eq:nl6})

For $s$ odd, antisymmetry under the exchange $2\leftrightarrow3$
allows only the following possible structures\[
R_{1}Q_{1}^{s-1}(R_{2}P_{2}+R_{3}P_{3})\,,\, Q_{1}^{s-1}(P_{2}S_{3}-P_{3}S_{2})\]
 The structure $R_{1}Q_{1}^{s-1}(R_{2}S_{2}+R_{3}S_{3})$ vanishes
on using eq. (\ref{eq:nl4}).

$\langle J_{1}J_{1}J_{0}\rangle$: The possible structures are \[
Q_{1}Q_{2}\,,\, P_{3}^{2}\,,\, R_{1}R_{2}P_{3}\,,\, R_{1}R_{2}S_{3}\,,\, P_{3}S_{3}\,,\, Q_{1}Q_{2}T\,,\, P_{3}^{2}T\]

$\langle J_{1}J_{1}J_{1}\rangle$: Note that all the parity even structures
that can occur in $\langle J_{1}J_{1}J_{1}\rangle$ are those that
are present in the non-linear relation eq.(\ref{eq:nl1}) but all
these structures are antisymmetric under the exchange of any two currents
whereas this correlator is symmetric under the same exchange. Hence
the parity even part of $\langle J_{1}J_{1}J_{1}\rangle$ vanishes.
For the same reason no possible parity odd structures can occur either.
Thus $\langle J_{1}J_{1}J_{1}\rangle$ vanishes in general.

$\langle J_{\frac{3}{2}}J_{\frac{1}{2}}J_{0}\rangle$: Here the possible
structures are \[
Q_{1}P_{3}\,,\, R_{1}R_{2}Q_{1}\,,\, Q_{1}S_{3}\,,\, Q_{1}P_{3}T\]

$\langle J_{\frac{3}{2}}J_{\frac{1}{2}}J_{\frac{1}{2}}\rangle$: The linearly independent structures are \[R_{1}Q_{1}P_{1}\,,\,R_{1}P_{2}P_{3}\,,\,Q_{1}(R_{2}P_{2}+R_{3}P_{3})\,,\,R_{1}Q_{1}S_{1} \]
Two other possible fermionic parity odd structures can be eliminated using eqs. (\ref{eq:nl4},\ref{eq:nl5})

$\langle J_{\frac{3}{2}}J_{\frac{1}{2}}J_{1}\rangle$: After eliminating some structures using the relations in sec. (7.2) we get the following linearly independent structures:
\begin{eqnarray*}
Q_{1}Q_{2}P_{2}\,,\,Q_{1}P_{1}P_{3}\,,\,P_{3}^{2}P_{2}\,,\,R_{1}R_{2}Q_{1}P_{1}\,,\,R_{1}R_{2}P_{2}P_{3}\,,\,R_{3}R_{1}Q_{1}Q_{2},\\
Q_{1}P_{1}S_{3}\,,\,Q_{1}P_{3}S_{1}\,,\,P_{2}P_{3}S_{3}\,,\,R_{1}R_{2}P_{2}S_{3}\,,\,Q_{1}Q_{2}P_{2}T\,,\,Q_{1}P_{1}P_{3}T\,,\,P_{3}^{2}P_{2}T\end{eqnarray*} 

$\langle J_{\frac{3}{2}}J_{\frac{3}{2}}J_{\frac{3}{2}}\rangle$: \begin{eqnarray*}
Q_{1}Q_{2}Q_{3}\sum_{i}R_{i}P_{i}\,,\,\,\,\,\sum_{cyclic}R_{1}Q_{2}Q_{3}P_{2}P_{3}\,,\,\,\,\,\sum_{i}R_{i}Q_{i}P_{i}^{3}\,,\,\,\,\, P_{1}P_{2}P_{3}\sum_{i}R_{i}P_{i}\,,\\
\sum_{i}R_{i}Q_{i}P_{i}^{2}S_{i}\end{eqnarray*}
 The structure $\sum_{cyc}R_{1}P_{1}(P_{2}^{2}Q_{2}+P_{3}^{2}Q_{3})$
can, by using the non-linear identity eq.(\ref{eq:nl1}), be expressed
in terms of the above structures and hence need not be included. The
structure $\sum_{cyclic}R_{1}Q_{2}Q_{3}(P_{2}S_{3}+P_{3}S_{2})$ vanishes
on using eqs. (\ref{eq:nl5},\ref{eq:nl4})

$\langle J_{2}J_{1}J_{1}\rangle$: The possible linearly independent structures are\begin{eqnarray*}
\,\,\,\,\,\,\,\,\,\,\,\,\,\,\,\,\,\,\, Q_{1}^{2}Q_{2}Q_{3}\,,\, Q_{1}^{2}P_{1}^{2}\,,\, Q_{1}P_{1}P_{2}P_{3}\,,\, P_{2}^{2}P_{3}^{2}\,,\\
R_{2}R_{3}P_{1}Q_{1}^{2}\,,\, R_{2}R_{3}P_{2}P_{3}Q_{1}\,,\\
Q_{1}Q_{2}P_{2}S_{2}+Q_{1}Q_{3}P_{3}S_{3}\,,\, P_{2}^{2}P_{3}S_{3}+P_{3}^{2}P_{2}S_{2}\,,\\
R_{1}R_{2}P_{2}^{2}S_{3}+R_{3}R_{1}P_{3}^{2}S_{2}\,,\\
Q_{1}^{2}Q_{2}Q_{3}T\,,\, Q_{1}^{2}P_{1}^{2}T\,,\, Q_{1}P_{1}P_{2}P_{3}T\,,\, P_{2}^{2}P_{3}^{2}T\end{eqnarray*}
Other structures are possible, but can be written in
terms of the other structures listed above by using the relations
in section \ref{relns}.

$\langle J_{3}J_{1}J_{1}\rangle$: As before, after eliminating some
structures which are antisymmetric under the exchange $2\leftrightarrow3$
we are left with the following linearly independent basis for $\langle J_{3}J_{1}J_{1}\rangle$
:\begin{eqnarray*}
Q_{1}^{2}(P_{2}^{2}Q_{2}-P_{3}^{2}Q_{3})\,,\\
Q_{1}^{2}(R_{1}R_{2}P_{1}P_{2}-R_{3}R_{1}P_{3}P_{1})\,,\, Q_{1}(R_{1}R_{2}P_{2}^{2}P_{3}-R_{3}R_{1}P_{3}^{2}P_{2})\,,\\
Q_{1}^{2}(P_{2}Q_{2}S_{2}-P_{3}Q_{3}S_{3})\,,\, Q_{1}(P_{3}^{2}P_{2}S_{2}-P_{2}^{2}P_{3}S_{3})\,,\\
Q_{1}(R_{1}R_{2}P_{2}^{2}S_{3}-R_{3}R_{1}P_{3}^{2}S_{2})\,,\,Q_{1}^{2}(P_{2}^{2}Q_{2}-P_{3}^{2}Q_{3})T\end{eqnarray*}

Again, linearly dependent structures have been eliminated using the relations of section \ref{relns}.

$\langle J_{4}J_{1}J_{1}\rangle$: The structures that occur here
are the same as $Q_{1}^{2}$ times the structures in $\langle J_{2}J_{1}J_{1}\rangle$. 

$\langle J_{s}J_{1}J_{1}\rangle$: For $s$ even this again equals
$Q_{1}^{s-2}\langle J_{2}J_{1}J_{1}\rangle$ (this was noted, for
the non-supersymmetric case, in ref. \cite{Giombi:2011rz}- it continues to hold
in our case). For $s$ odd and greater than three this correlator
equals $Q_{1}^{s-2}\langle J_{3}J_{1}J_{1}\rangle$. Thus the number
of possible tensor structures in $\langle J_{s}J_{1}J_{1}\rangle$
does not increase with $s$.

$\langle J_{2}J_{2}J_{2}\rangle$: The following are the possible
independent invariant structures\begin{eqnarray*}
Q_{1}^{2}Q_{2}^{2}Q_{3}^{2}\,,\,\,\,\,\, P_{1}^{2}P_{2}^{2}P_{3}^{2}\,,\,\,\,\,\, Q_{1}Q_{2}Q_{3}P_{1}P_{2}P_{3}\,,\,\,\,\,\,\sum_{i}Q_{i}^{2}P_{i}^{4}\,,\\
Q_{1}Q_{2}Q_{3}\sum_{cyclic}Q_{3}P_{3}R_{1}R_{2}\,,\,\,\, P_{1}P_{2}P_{3}\sum_{cyclic}Q_{3}P_{3}R_{1}R_{2}\,,\\
P_{1}P_{2}P_{3}\sum_{cyclic}P_{1}P_{2}S_{3}\,\,,\,\,\,\,\,\sum_{i}Q_{i}^{2}P_{i}^{3}S_{i}\,,\,\,\,\,\,\\
Q_{1}Q_{2}Q_{3}\sum_{cyclic}Q_{3}S_{3}R_{1}R_{2}\,,\,\,\,\,\, P_{1}P_{2}P_{3}\sum_{cyclic}Q_{3}S_{3}R_{1}R_{2}\,,\\
Q_{1}^{2}Q_{2}^{2}Q_{3}^{2}T\,,\,\,\,\,\, P_{1}^{2}P_{2}^{2}P_{3}^{2}T\,,\,\,\,\,\, Q_{1}Q_{2}Q_{3}P_{1}P_{2}P_{3}T\,,\,\,\,\,\,\sum_{i}Q_{i}^{2}P_{i}^{4}T\end{eqnarray*}
Many other linearly dependent structures have been eliminated using the relations in sec. (7.2).

As is evident, the number of invariant structures needed to construct
the 3-point correlator increases rapidly as the spins of the operators
increase and we will not consider more examples. 

It is clear from the above examples that the general structure of
the 3-point function is the following:\begin{equation}
\langle J_{s_{1}}J_{s_{2}}J_{s_{3}}\rangle=\frac{1}{\tilde{X}_{12}^{m_{123}}\tilde{X}_{23}^{m_{231}}\tilde{X}_{31}^{m_{312}}}\sum_{n}\mathcal{F}_{n}(P_{i},\, Q_{i},\, R_{i},\, S_{i},\, T)\end{equation}
where $m_{ijk}\equiv (\Delta_{i}-s_{i})+(\Delta_{j}-s_{j})-(\Delta_{k}-s_{k})$ and the sum is over all the independent invariant structures $\mathcal{F}_{n}$,
each of homogeneity $\lambda_{1}^{2s_{1}}\lambda_{2}^{2s_{2}}\lambda_{3}^{2s_{3}}$.
Since the 3-point function is linear in the parity odd invariants
and linear or bilinear in the $R$'s (either $R_{i}$ or $R_{j}R_{k}\,,\, j\ne k$),
we have the following structure for $\mathcal{F}_{n}$:\begin{eqnarray*}
\mathcal{F}_{n}=F_{n}^{(1)}(P_{i},Q_{i})+a_{n}^{(1)}F_{n}^{(1)}(P_{i},Q_{i})T+a_{n}^{(2)}F_{n}^{(2)}(P_{i},Q_{i})S_{i}+a_{n}^{(3)}F_{n}^{(3)}(P_{i},Q_{i})R_{i}\\
+a_{n}^{(4)}F_{n}^{(4)}(P_{i},Q_{i})R_{i}S_{j}+a_{n}^{(5)}F_{n}^{(5)}(P_{i},Q_{i})R_{j}R_{k}+a_{n}^{(6)}F_{n}^{(6)}(P_{i},Q_{i})R_{j}R_{k}S_{l}\end{eqnarray*}
 Here each $F_{n}^{(a)}(P_{i},Q_{i})$ is a monomial in $P$'s and
$Q$'s such that each term on the r.h.s above has homogeneity $\lambda_{1}^{2s_{1}}\lambda_{2}^{2s_{2}}\lambda_{3}^{2s_{3}}$.%
\footnote{The six $F_{n}^{(a)}(P_{i},Q_{i})$ are not independent functions.
$F_{n}^{(2)},\, F_{n}^{(4)},\, F_{n}^{(6)}$can be obtained from $F_{n}^{(1)},\, F_{n}^{(3)},\, F_{n}^{(5)}$,
respectively, by replacing a $P_{i}^{p}$ in the latter by $P_{i}^{p-1}S_{i}$ (suitably
(anti-)symmetrized if some spins are equal in the 3-point function)
}

\subsubsection{Three point functions of conserved currents}\label{3pfcc}
We have so far considered the constraints on the structure of the three-point functions of higher spin operators arising due to superconformal invariance alone. We will now see how the structure is further constrained by current conservation, i.e, when the operators are actually conserved higher spin currents. In this section we present evidence for the claim that the three point function of the conserved higher spin currents in ${\cal N}=1$ superconformal field theory consists
of two linearly independent parts, i.e.,
\begin{equation}
 \langle J_{s_{1}}J_{s_{2}}J_{s_{3}}\rangle = \frac{1}{{\tilde X}_{12}{\tilde X}_{23}{\tilde X}_{31}}\left(a \langle J_{s_{1}}J_{s_{2}}J_{s_{3}}\rangle_{\mbox{even}} 
                                              + b \langle J_{s_{1}}J_{s_{2}}J_{s_{3}}\rangle_{\mbox{odd}}\right)
\end{equation}
where $a$ and $b$ are independent constants, and the `even' structure arises from free field theory.

The procedure, quite similar to that used by \cite{Giombi:2011rz}, is as follows. For any particular three point function we first  consider the linearly independent basis of monomial structures (listed in section \ref{indep}) and take an arbitrary linear combination of these structures.
\begin{equation} \langle J_{s_{1}}J_{s_{2}}J_{s_{3}}\rangle =\frac{1}{{\tilde X}_{12}{\tilde X}_{23}{\tilde X}_{31}}\sum_{n}a_{n}\mathcal{F}_{n}\end{equation}

Current conservation $D_{\alpha_{1}}J^{\alpha_{1}\alpha_{2}.....\alpha_{2s}}=0$ is tantamount to the following equation on the contracted current $J_{s}(x,\lambda)$ :
\begin{equation}
D_{\alpha}\frac{\partial}{\partial\lambda _{\alpha}}J_{s}=0\end{equation}

Thus the equation
\begin{equation}D_{i}\frac{\partial}{\partial\lambda _{i}} \langle J_{s_{1}}J_{s_{2}}J_{s_{3}}\rangle=0\end{equation}
for each $i=1,2,3$ gives additional constraints in the form of linear equations in the $a_{n}$'s- some of these constants can thus be determined. The algebraic manipulations get quite unwieldy- we used superconformal invariance to set some co-ordinates to particular values and took recourse to Mathematica. The results obtained are given below (the known $\tilde{X}_{ij}$ dependent factors in the denominator
are not listed below):
\begin{table}[h]
\centering
\begin{tabular}{| c | c | c |}
\hline
Three-pt function & Even & Odd \\
\hline
$\langle J_{\half}J_{\half} J_0\rangle$ & $P_3 - \frac{i}2 R_1 R_2 $  & $S_3 - \frac{i}2 P_3T$\\
\hline
$\langle J_1J_\half J_0 \rangle$ & $P_3 R_1 + \half Q_1 R_2$ & 0\\
\hline
$\langle J_1 J_1 J_0 \rangle$ & $\half Q_1Q_2 + P_3^2 -i R_1 R_2P_3$ & $S_3P_3 + \frac{i}2(S_3 R_1 R_2 - Q_1 Q_2 T)$\\
\hline
$\langle J_{\frac{3}{2}}J_{\frac{1}{2}}J_{0}\rangle$ & $P_{3}Q_{1}-\frac{i}{2}Q_{1}R_{1}R_{2}$ & $Q_1S_3-iQ_1P_3T$\\
\hline
$\langle J_{\frac{3}{2}}J_{\frac{1}{2}}J_\half\rangle$ & $Q_1R_1P_1 + Q_1(R_2P_2 + R_3P_3) + 2 R_1P_2P_3$ & 0\\
\hline
$\langle J_2J_\half J_\half\rangle$ & $Q_1^2 P_1 -4 Q_1 P_2P_3 - \frac{5i}2 R_2R_3Q_1^2$ &$Q_1(P_2S_3+P_3S_2)$ \\
& & $+ \frac{i}2(Q_1^2P_1-3Q_1P_2P_3)T$\\
\hline
% $\langle J_{2}J_{1}J_{1}\rangle$ & $Q_1^2 P_1^2+Q_1^2Q_2Q_3 + \frac{8}5 P_2^2P_3^2 - i R_2 R_3 P_1 Q_1^2$ & \\
%                                   & $+ 4i R_2 R_3 P_2 P_3Q_1$ &\\ 
% \hline
\end{tabular}
\caption{Explicit examples of conserved three-point functions.}\label{egtab}
\end{table}

Using expression (\ref{expression1}) for the currents in the $\mathcal{N}=1$ 
free theory, some 3-point functions were explicitly evaluated (again using Mathematica, 
s the computations get quite cumbersome beyond a few lower spin examples). It must be 
emphasized that the (tabulated) even structures obtained above match 
with the expressions obtained from free field theory (upto overall constants). 
We thus have some evidence for the claim that the three-point function of conserved 
currents has a parity even part (generated by a free field theory) and a parity 
odd piece.

\section{Summary and Outlook}
In this paper we have embarked on the study of superconformal Chern-Simons matter 
theories in an on-shell superspace formalism. To conclude we summarize the main results 
obtained in this paper below.
\begin{itemize}
 \item An explicit construction of higher spin conserved supercurrents in terms of 
 higher spin component currents in section \ref{genstrcur}.
 \item An explicit construction of higher spin conserved supermultiplets in terms of
 on shell elementary superfields in free superconformal field theories in section \ref{ffc}.
 \item A decomposition of the state content of single trace operators in large $N$ 
 vector Chern-Simons superconformal theories into multiplets of the superconformal 
 algebra in the theories with ${\cal N}=1,2,3,4,6$ superconformal symmetry in appendix \ref{repth}.
 \item Determination of the form of two point functions of conserved higher spin 
 supercurrents, and the explicit computation of these 2 point functions in free 
 theories in section \ref{2pf}. 
 \item Classification of superconformal invariants formed out of 3 polarization 
 spinors and 3 superspace insertion points (following \cite{Park:1999cw}) and use 
 thereof to constrain 3 point functions of higher spin operators in 3d superconformal 
 field theories in section \ref{N13pfinv}.
 \item A conjecture - and evidence - that there are exactly two structure allowed 
 in the 3 point functions of the conserved higher spin currents for ${\cal N}=1$ in section \ref{3pfcc}. 
 \item The superspace structure of higher spin symmetry breaking on 
 adding interactions to large $N$ gauge theories in section \ref{wbc}. 
\end{itemize}

One of the main motivations for the study embarked upon in this paper is to perform a 
Maldacena-Zhiboedov type study of superconformal Chern-Simons vector matter theories. As shown 
in the section \eqref{wbc}, the structure of terms violating higher spin current conservation 
is much more constrained in superconformal case as compared to the conformal case suggesting 
that higher spin correlators in superconformal case must be more severely constrained. For 
this purpose it will be useful to extend the analysis of three-point functions presented here 
for ${\cal N}=1$ case to extended supersymmetry. Besides describing a variety of 
renormalization group fixed points in 3 dimensions, theories of this type are also expected 
to be holographic duals to supersymmetric higher spin Vasiliev theories in 4 dimensions. It 
may also be worth extending this formalism for 4 and higher point functions by using 
polarization spinor techniques, perhaps together with the embedding formalism, in view 
of implementing the (super)conformal bootstrap for higher spin operators.

\acknowledgments
We thank Prof. Shiraz Minwalla for suggesting this problem to us 
and for his guidance and helpful suggestions during the course of this project. 
We also thank R. Loganayagam and S. Prakash for useful discussions. We 
acknowledge the use of Matthew Headrick's {\it grassmann} Mathematica 
package for doing computations with fermionic variables. AAN  thanks 
the Dept. of Theoretical Physics, TIFR, Mumbai and also the Crete Centre for 
Theoretical Physics, Heraklion (Erasmus IP in Non-perturbative QFT) for 
hospitality while this work was in progress. AAN would also like to thank 
Trinity College, Cambridge for a Rouse Ball Travelling Studentship and the 
Cambridge Commonwealth Trust for partial financial support. TS would like to 
thank ICTP, Trieste for hospitality while this work was in progress. Finally, 
TS and VU would like acknowledge our debt to the people of India for their generous 
and steady support to research in the basic sciences.

\appendix
\appendixpage
\addappheadtotoc

\section{Conventions} \label{con}
\subsection{Spacetime spinors}\label{con1}
The Lorentz group in $D=3$ is $SL(2,\mathbb{R})$ (see, for instance, the appendix of \cite{Dumitrescu:2011iu})
and we can impose the Majorana condition
on spinors, i.e., the fundamental representation is a real two component spinor $\psi_\alpha = \psi^*_\alpha$
($\alpha = 1,2$). The metric signature is mostly plus. $D=3$ 
superconformal theories with ${\cal N}$ extended supersymmetry posses an $SO({\cal N})$
$R$-symmetry which is part of the superconformal algebra,
whose generators are real antisymmetric matrices $I^{ab}$, where $a,b$ are the vector indices of
$SO({\cal N})$.
The supercharges carry a vector $R$-symmetry index, $Q_\alpha^a$, as do the superconformal generators $S_\alpha^a$.

In $D=3$ we can choose a real basis for the $\gamma$ matrices
\be
(\gamma_\mu)_\alpha^{~\beta} \equiv (i\sigma^2,\sigma^1,\sigma^3) = \left(\left(\begin{array}{cc} \mbox{  }0 & \mbox{  }1  \\ -1 & \mbox{ }0 \end{array}\right),
\left(\begin{array}{cc} 0 & 1  \\ \mbox{ }1 & \mbox{ }0 \end{array}\right),\left(\begin{array}{cc} 1 & \mbox{    }0  \\ 0 & -1 \end{array}\right)\right)
\ee
Gamma matrices with both indices up (or down) are symmetric
\begin{equation}
\begin{split}
 (\gamma_\mu)_{\alpha\beta} \equiv (\mathds{1},\sigma^3,-\sigma^1) &~~~~~~~ (\gamma_\mu)^{\alpha\beta} \equiv (\mathds{1},-\sigma^3,\sigma^1)
\end{split}
\end{equation}
The antisymmetric $\epsilon$ symbol is $\epsilon^{12} = -1 = \epsilon_{21}.$
It satisfies
\be
\begin{split}
\epsilon \gamma^\mu \epsilon^{-1} = -(\gamma^\mu)^T\\
\epsilon \Sigma^{\mu\nu} \epsilon^{-1} = -(\Sigma^{\mu\nu})^T
\end{split}
\ee
where $\Sigma^{\mu\nu} = -\frac{i}{4}[\gamma^\mu,\gamma^\nu]$ are the Lorentz generators.
The charge conjugation matrix $C$ can be chosen to be the identity, which we take to be 
\be
\begin{split}
-\epsilon\gamma^0 = C^{-1} &~~~~~  \gamma^0\epsilon^{-1} = C
\end{split}
\ee
$C^{\alpha\beta}$ denotes the inverse of $C_{\alpha\beta}$.
Spinors transform as follows
$$\psi^\prime_\alpha \rightarrow -(\Sigma_{\mu\nu})_\alpha^{~\beta}\psi_\beta.$$
Spinors are naturally taken to have index structure down, i.e., $\psi_\alpha$.

The raising and lowering conventions are
\be
\begin{split}
\psi^\beta = \epsilon^{\beta\alpha}\psi_\alpha\\
\psi_\alpha = \epsilon_{\alpha\beta}\psi^\beta
\end{split}
\ee
There is now only one way to suppress contracted spinor indices,
$$\psi\chi = \psi^\alpha\chi_\alpha,$$
and this leads to a sign when performing Hermitian conjugation
$$(\psi\chi)^* = -\chi^*\psi^*.$$
The $\gamma$ matrices satisfy
\be
(\gamma_\mu\gamma_\nu)_\alpha^{~\beta} =  \eta_{\mu\nu}\delta_{\alpha}^{~\beta} + \epsilon_{\mu\nu\rho}(\gamma^\rho)_\alpha^{~\beta}
\ee
where $\epsilon_{\mu\nu\rho}$ is the Levi-Civita symbol, and we set $\epsilon_{012} = 1$ ($\epsilon^{012} = -1$). The
superconformal algebra is given below:
\begin{equation}
\begin{split}
[ M_{\mu\nu}, M_{\rho\lambda} ]&= i\left( \eta_{\mu\rho}M_{\nu\lambda}- \eta_{\nu\rho}M_{\mu\lambda}- 
                                          \eta_{\mu\lambda}M_{\nu\rho}+ \eta_{\nu\lambda}M_{\mu\rho} \right),   \\
[ M_{\mu\nu}, P_\lambda ]&= i(\eta_{\mu\lambda}P_\nu- \eta_{\nu\lambda}P_\mu),  \\
[ M_{\mu\nu}, K_\lambda ]&= i(\eta_{\mu\lambda}K_\nu- \eta_{\nu\lambda}K_\mu),  \\
[ D, P_\mu ]&= iP_{\mu} ~,~~~ [ D, K_\mu ]= -iK_\mu, \\
[ P_\mu, K_\nu ]&= 2i(\eta_{\mu\nu} D- M_{\mu\nu}),  \\
[ I_{ab}, I_{cd} ] &= i\left(\delta_{ac}I_{bd} - \delta_{bc}I_{ad} - \delta_{ad}I_{bc} + \delta_{bd}I_{ac}\right),\\
% [ I_{ab}, M_{\mu\nu} ] &= [ I_{ab}, P_\mu]=0, \\
\{ Q_\alpha^a, Q_\beta^b \}&= (\gamma^{\mu})_{\alpha\beta}P_\mu\delta^{ab}, \\
% \{ D_\alpha, D_\beta \}&= -(\gamma^{\mu})_{\alpha\beta}P_\mu, \\
[I_{ab}, Q^\alpha_c] &= i(\delta_{ac}Q^\alpha_b - \delta_{bc}Q^\alpha_a) , \\
\{ S_\alpha^a, S_\beta^b \}&= (\gamma^{\mu})_{\alpha\beta}K_\mu\delta^{ab}, \\
[I_{ab}, S^\alpha_c] &= i(\delta_{ac}S^\alpha_b - \delta_{bc}S^\alpha_a) , \\
[ K_\mu, Q_\alpha^a ]&= i(\gamma_\mu)_{\alpha}^{~\beta}S_{\beta}^a, \\
[ P_\mu, S_\alpha^a ]&= i(\gamma_\mu)_{\alpha}^{~\beta}Q_{\beta}^a, \\
[ D, Q_\alpha^a ]&= \frac{i}{2} Q_\alpha^a ~,~~~ [ D, S_\alpha^a ]= -\frac{i}{2} S_\alpha^a, \\
[M_{\mu\nu}, Q_\alpha^a]&= -(\Sigma_{\mu\nu})_\alpha^{~\beta}Q_\beta^a, \\
[M_{\mu\nu}, S_\alpha^a]&= -(\Sigma_{\mu\nu})_\alpha^{~\beta}S_\beta^a, \\
\{ Q^a_\alpha, S^b_\beta \}&= \left(\epsilon_{\beta\alpha}D - \half\epsilon_{\mu\nu\rho}(\gamma^\rho)_{\alpha\beta}M^{\mu\nu}\right)\delta^{ab} + \epsilon_{\beta\alpha}I^{ab}.
\end{split}
\label{N1lag}
\end{equation}
All other (anti)-commutators vanish.

\subsection{$R$-symmetry}\label{Rsymconv}

\subsubsection{$SO(3)$}\label{con12}

Gamma matrices are chosen to be the sigma matrices
\begin{equation}\label{so3gama}
(\sig^a)_i^{~j}=\left( \left( \begin{tabular}{cc}
               0 & 1 \\
               1 & 0 \\
              \end{tabular} \right),
              \left( \begin{tabular}{cc}
               0 & $-i$ \\
               $i$ & 0 \\
              \end{tabular} \right),
              \left( \begin{tabular}{cc}
               1 & 0 \\
               0 & $-1$ \\
              \end{tabular} \right) \right).
\end{equation}
Indices are raised and lowered by $\e^{12}=-1=-\e_{12}$. Note that $\sigma$ matrices with both lower 
or both upper indices are symmetric.

The following identities are useful
\be\begin{split}\label{so3iden}
\e^{ij}\phi^k &+ \e^{jk}\phi^i + \e^{ki}\phi^j = 0, \\
\epsilon^{ij} \epsilon_{kl} &= \delta_l^i \delta_k^j- \delta _k^i \delta_l^j, \\ 
\epsilon_{ij} \epsilon_{kl} &= \epsilon_{ik} \epsilon_{jl} - \epsilon_{il} \epsilon_{jk}, 
                          \quad (\text{same for upper indices})   \\
(\sigma^a)_i^{~j}  (\sigma^a)_k^{~l} &= 2 \delta _i^l \delta_k^j- \delta_i^j \delta_k^l \\
(\sigma^a)_{ij}  (\sigma^a)_{kl} &= -(2 \e_{il} \e_{jk}+ \e_{ij} \e_{kl}) 
                                  = -(\e_{ik}\e_{jl} + \e_{il}\e_{jk}) \\
\end{split}
\ee

\subsubsection{$SO(4)$}\label{con13}
Gamma matrices are chosen to be 
\be\begin{split}\label{so4gama}
\Gamma^a &= \left( \begin{tabular}{ll}
           0 & $\sigma^a$ \\
           $\bar\sigma^a$ & 0 \\
           \end{tabular} \right)  ~\text{for $a$= 1,2\ldots 4} \\
\text{where} ~ 
(\sigma^a)_i^{~\tilde{i}} &= (\sigma^1,\sigma^2,\sigma^3,i\mathds{1}_2), \quad
(\bar\sigma^a)_{\tilde{i}}^{~i} = (\sigma^1,\sigma^2,\sigma^3,-i\mathds{1}_2). 
\end{split}
\ee
Indices are raised and lowered by $\e^{12}=-\e_{12}=-1=\tilde{\e}^{12}=-\tilde{\e}_{12}$.
With these definitions, the following identities would be useful.
\be\begin{split}
(\bar\sigma^a)^{\tilde{i}i} &=(\bar\sigma^{aT})^{\tilde{i}i} \quad \big((\bar\sigma^a)^T=-\e\sigma^a\tilde{\e}^{-1}\big), \\
(\sigma^a)^{~\tilde{i}}_i (\bar\sigma^a)_{\tilde{j}}^{~j} &= 2\D^{\tilde{i}}_{\tilde{j}} \D^j_i, \\
(\sigma^a)^{i\tilde{i}} (\bar\sigma^a)^{\tilde{j}j} &= -2\e^{ij}\e^{\tilde{i}\tilde{j}}, 
\quad (\sigma^a)_{i\tilde{i}} (\bar\sigma^a)_{\tilde{j}j} = -2\e_{ij}\e_{\tilde{i}\tilde{j}}. \\
\end{split}
\ee

\subsubsection{$SO(6)$}\label{con14}
We choose the gamma matrices to be 
\be\begin{split}\label{so6gama}
\Gamma^a &= \left( \begin{tabular}{ll}
           0 & $\gamma^a$ \\
           $\bar\gamma^a$ & 0 \\
           \end{tabular} \right)  ~\text{for~} a= 1,2\ldots 6 \\
\text{where} ~ 
\gamma^a &= (\gamma^1,\gamma^2,\gamma^3,\gamma^4,\gamma^5,i\mathds{1}_4), ~
\bar\gamma^a = (\gamma^1,\gamma^2,\gamma^3,\gamma^4,\gamma^5,-i\mathds{1}_4), ~
\gamma^5=\gamma^1\gamma^2\gamma^3\gamma^4. \\
\text{and}~
\gamma^i &= \left( \begin{tabular}{ll}
          0 & $\sigma^i$ \\
          $\bar\sigma^i$ & 0 \\
          \end{tabular} \right)
          \text{with}~ \sigma^i=(\sigma^1,\sigma^2,\sigma^3,i\mathds{1}_2), 
                       \bar\sigma^i= (\sigma^1,\sigma^2,\sigma^3,-i\mathds{1}_2) ~
           \text{for~} i=1 \ldots 4
\end{split}
\ee
In these basis we the `chirality' projection matrix is diagonal and is given by
\be 
\Gamma^{7}= -i\Gamma^0\Gamma^1\Gamma^2\Gamma^3\Gamma^4\Gamma^5= 
\left( \begin{tabular}{cc}
$I_4$ & 0 \\
0 & -$I_4$ \\
\end{tabular} \right) 
\ee
The charge conjugation matrix is 
\be
C= \Gamma^0\Gamma^2\Gamma^4= \left( \begin{tabular}{ll}
                             0 & $c$ \\
                             -$c$ & 0 \\ 
                            \end{tabular} \right)
\text{with}~ c= i\gamma^2\gamma^4 
\ee
which satisfies 
\be\begin{split}\label{ccprop}
C^* &= C^{-1} = -C, \quad (\Gamma^a)^* = C^{-1}\Gamma^a C \\
\Rightarrow \quad c &= -c^* = c^{-1}, \quad (\bar{\gamma}^a)^* = -c^{-1}\gamma^a c \\
\text{In index notation:} \quad (\bar{\gamma}^{a~j}_{~i})^* &= (\bar{\gamma}^{a*})^i_{~j} 
                                                             = -c^{ik} (\gamma^a)_{k}^{~l} c_{lj} 
                                                             = c^{ik}  c_{jl} (\gamma^a)_{k}^{~l}
\end{split}
\ee

Indices are raised and lowered with using the charge conjugation matrix $C$ for $\Gamma^a$ and 
$c$ for $\gamma^a$. With both indices up or down the $\gamma$ matrices are 
antisymmetric\footnote{This should be the case as the vector of $SO(6)$ is 
$(4\times4)_{\mbox{antisym}}$ of $SU(4)$.}. The last equation in \eqref{ccprop} implies the following \
useful properties for the generators  
Let us define 
$$ \gamma^{ab} = \gamma^a\bar{\gamma}^b- \gamma^b\bar{\gamma}^a, \quad 
\bar{\gamma}^{ab} = \bar{\gamma}^a\gamma^b- \bar{\gamma}^b\gamma^a, $$
then we have following useful relations 
\be\begin{split}\label{so6genprop}
\gamma^{ab\dagger} &= -\gamma^{ab}, \quad 
\bar{\gamma}^{ab\dagger} = -\bar{\gamma}^{ab}, \\
(\bar{\gamma}^{ab*})^i_{~j} &= (c^{-1}\gamma^{ab}c)^i_{~j}, \quad 
(\bar{\gamma}^{ab})_i^{~j} = -(c^{-1}\gamma^{ab}c)^j_{~i}.\\
\end{split}
\ee
The first line says that the generators of $SO(6)$ transformation are Hermitian\footnote{The 
generator of $SO(6)$ acting on chiral and antichiral transformation are 
respectively $-\frac{i}{4}\gamma^{ab}$ and $-\frac{i}{4}\bar{\gamma}^{ab}$} while the 
two equation in the second line follows from \eqref{ccprop}.

The following identities are useful\footnote{Note that representation theory ($SU(4)$) wise $C$ shouldn't be used 
to raise or lower indices as it is not an invariant tensor of $SU(4)$. Only $\e^{ijkl}$ and 
$\e_{ijkl}$(which are specific combinations of product of c's) can be used to raise or lower 
$SU(4)$ indices. we will explicitly see that all the $SU(4)$ tensor equations can be written 
using just $\e$ tensors.}:
\be\begin{split}\label{so6iden}
\bar{\gamma}^a_{ij} &= \gamma^a_{ij} + 2\delta^{a0}c_{ij},
\quad (\bar{\gamma}^a)_i^{~j} = (\gamma^a)_i^{~j} - 2\delta^{a0}\delta_i^j,  \\
\gamma^a_{ij}\gamma^a_{kl} &= -2\epsilon_{ijkl}= 2(c_{ik}c_{jl}-c_{il}c_{jk}-c_{ij}c_{kl}), \\
\gamma^a_{ij}\bar{\gamma^a}_{kl} &= -2\epsilon_{ijkl}+ 2c_{ij}c_{kl}= 2(c_{ik}c_{jl}-c_{il}c_{jk}), \\
(\gamma^a)^{ij}(\bar{\gamma^a})_{kl} &= 2\delta^i_k\delta^j_l -2\delta^i_l\delta^j_k, \\
(\gamma^{ab})_i^j(\gamma^{ab})_k^l &= -32\delta_i^l\delta_k^j + 8\delta_i^j\delta_k^l, \\
\end{split}
\ee

\subsection{Useful relations}
Some useful relations and identities are given below
\begin{eqnarray}
% \theta\theta &=& -2\theta^1\theta^2 = 2 \theta_1 \theta_2 \\
\epsilon^{\alpha\beta}\frac{\partial}{\partial\theta^{a\beta}} = -\frac{\partial}{\partial\theta^a_\alpha} \\
(\gamma^\mu)_{\alpha}^{~\beta}(\gamma_\mu)_{\sigma}^{~\rho} = 2 \delta^{~\rho}_\alpha \delta_\sigma^{~\beta} -\delta_\alpha^{~\beta}\delta_\sigma^{~\rho}\\ 
\theta_\alpha \theta_\beta = \half \epsilon_{\alpha\beta} \theta\theta, \,\,\,\,\, 
\theta^\alpha \theta^\beta = -\half \epsilon^{\alpha\beta} \theta\theta\\
\theta_{1\alpha}\theta_{2}^{\beta}+\theta_{2\alpha}\theta_{1}^{\beta}+(\theta_{1}\theta_{2})\delta_{\alpha}^{\beta}=0\\
X^{2}\equiv X_{\alpha}^{\,\,\beta}X_{\beta}^{\,\,\alpha}=2x_{\mu}x^{\mu}\equiv2x^{2}\,,\,\,\,\,\,\,\, X_{\alpha}^{\,\,\beta}X_{\beta}^{\,\,\gamma}=x^{2}\delta_{\alpha}^{\gamma}=\frac{X^{2}}{2}\delta_{\alpha}^{\gamma}\\
D_{1\alpha}\tilde{X}_{12}^{-\Delta}=i\Delta(\tilde{X}_{12})_{\alpha}^{\,\,\beta}(\theta_{12})_{\beta}\\
D_{1\alpha}(\tilde{X}_{12})_{\beta}^{\,\,\gamma}=-i\delta_{\alpha}^{\gamma}(\theta_{12})_{\beta}+\frac{i}{2}\delta_{\beta}^{\gamma}(\theta_{12})_{\alpha}\\
D_{1\alpha}(X_{12-})_{\beta}^{\,\,\gamma}=-i\delta_{\alpha}^{\gamma}\theta_{12\beta}\,,\,\,\,\,\, D_{1\alpha}(X_{12+})_{\beta}^{\,\,\gamma}=i\epsilon_{\alpha\beta}\theta_{12}^{\gamma}
%\theta_\alpha \chi_\beta - \theta_\beta \theta_\alpha &=& -\epsilon_{\alpha\beta} \theta\chi \\
\label{iden}
\end{eqnarray}

\section{Conformal spectrum of free scalars and fermions}\label{scalfermcd}
In this appendix we list the character decomposition of product of two short conformal representations 
into irreducible conformal representations for the particular cases of a complex scalar and 
complex fermions. Let us denote by $\chi(\phi)$ and $\chi(\psi)$ the conformal character of 
a free scalar field and a free fermion (in $D=3$) respectively. Then we have 
\begin{equation}
\begin{split}
\chi(\phi)\chi(\bar\phi) &= \frac{1}{(1-x)(1-xy)(1-xy^{-1})}\left( x+ \sum^{\infty}_{k=1} \chi_{(sh)}(k+1,k) \right) \\
\chi(\bar\phi)\chi(\psi)= \chi(\phi)\chi(\bar\psi) &= \frac{1}{(1-x)(1-xy)(1-xy^{-1})}\left( x^{\frac{3}{2}}\chi_\half(y) + 
                            \sum^{\infty}_{k=1} \chi_{(sh)}(k+\frac{3}{2},k+\half) \right) \\
\chi(\psi)\chi(\bar\psi) &= \frac{1}{(1-x)(1-xy)(1-xy^{-1})}\left( x^2 + \sum^{\infty}_{k=1} \chi_{(sh)}(k+1,k) \right) \\
\end{split}
\label{chrdecmp}
\end{equation}
where the $\chi_{(sh)}(j+1,j)$ denotes the character of a {\it short} conformal representation 
with spin $j$.

Let us consider a free ${\cal N}=1$ superconformal theory of a complex boson and a complex fermion 
transforming in $N$ of $SU(N)$ gauge group. The spectrum of gauge invariant single trace operators 
in theory is then just the sum of the operators 
represented in \eqref{chrdecmp}. Using the decomposition in \eqref{N1decmp}, the operators in 
\eqref{chrdecmp} are easily combined into the representation of ${\cal N}=1$ supermultiplets. 
These are given as follows \footnote{Here $(\Delta,j)_1$ denote ${\cal N}=1$ representation 
while $(\Delta,j)$ denotes a conformal representation.}
\begin{equation}\label{N1spctrm}
\left( \half,0 \right)_1 \oplus \sum_{k=1}^\infty \left( \frac{k}{2}+1,\frac{k}{2} \right)_1
\end{equation}
i.e. along with the special short representation with spin zero there are superconformal short 
representation for every positive half integer spin starting from spin $\half$. For 
convenience we list the decomposition of all short and long ${\cal N}=1$ superconformal 
representations below
\begin{equation}\begin{split}
(\Delta, j)_{1,long} &= (\Delta,j) \oplus \left(\Delta+\half, j-\half\right) \oplus 
                         \left(\Delta+\half, j+\half\right) \oplus (\Delta+1, j), \\
(j+1, j)_1 &= (j+1, j) \oplus \left(j+\frac{3}{2}, j+\half\right), \\
\left( \half, 0 \right)_1 &= \left( \half,0 \right) \oplus \left( 1, \half \right) \oplus 
                          \left( \frac{3}{2},0 \right).  \\
\end{split}
\label{N1decmp}
\end{equation}

\section{Superconformal spectrum of ${\cal N}=1,2,3,4,6$ theories}\label{repth} 
In this appendix we discuss the full single trace gauge invariant local operator spectrum 
of the free $U(N)$ superconformal Chern-Simons vector theories discussed in section \ref{fscftss}. 
In subsequent subsections here we present the full conformal primary spectrum, using the conformal 
grouping discussed in appendix \ref{scalfermcd}, and then group 
these conformal primaries into representations of superconformal algebra of the 
respective theory\footnote{Representation of superconformal algebra are labeled 
by the scaling dimension, spin and R-symmetry representation of the superconformal 
primary. see e.g. section (3) of \cite{Minwalla:2011ma} a summary of unitary representations 
of superconformal algebra in 2+1d.}.

\subsection{${\cal N}=1$}\label{n1spectrum}
The minimal field content of this theory consists of a complex scalar and a complex fermion. The conformal
content is easy to write down; there are both integer and half-integer spin currents in the theory. All of
these group into short superconformal multiplets of both integer and half integer spin. Thus, the superconformal
primary content of this theory is
\be\bigoplus_{j=0,\half,1,\ldots}^\infty (j+1,j)_{{\cal N}=1},\ee
where $(j+1,j)$ denotes a dimension $j+1$, spin $j$ short superconformal primary multiplet
which contains the conserved spin $j$
and spin $j+\half$ conformal primaries. There is no $R$-symmetry quantum number in this case.
The conformal content is
\be
\begin{split}
(j+1,j)_{{\cal N}=1} &\rightarrow (j+1,j) \oplus (j+\frac{3}2,j+\half) ~~~~~ j\neq0 \\
(1,0)_{{\cal N}=1} &\rightarrow (1,0) \oplus \left(\frac{3}2,\half\right) \oplus (2,0) ~~~~~ j=0.
\end{split}
\ee

\subsection{${\cal N}=2$}\label{N2spectrum}
The field content of the ${\cal N}=2$ theories is the same as that of ${\cal N}=1$; the difference
being that the spectrum of short superconformal multiplets consists only of integer spins. Thus,
we can write the spectrum of short superconformal primaries in these theories as
\be\bigoplus_{j=0,1,\ldots}^\infty (j+1,j,0)_{{\cal N}=2}.\ee\label{N2spectrum_1}
The conformal content for a spin $j$ ${\cal N}=2$ short superconformal primary in terms of ${\cal N}=1$ is
\be
(j+1,j,0)_{{\cal N}=2} \rightarrow  (j+1,j)_{{\cal N}=1} \oplus (j+\frac{3}2,j+\half)_{{\cal N}=1}
\ee
from which he conformal content can be read off as
\be
\begin{split}
(j+1,j,0)_{{\cal N}=2} &\rightarrow  (j+1,j,0) \oplus (j+\frac{3}2,j+\half,1) \oplus (j+\frac{3}2,j+\half,-1) \oplus (j+2,j+1,0) ~~ j\neq0\\
(1,0,0)_{{\cal N}=2} &\rightarrow (1,0,0) \oplus (\frac{3}2,\half,1) \oplus (2,0,0) \oplus (\threehalf,\half,-1) \oplus (2,1,0) ~~~~ j=0.
\end{split}
\ee
where the third quantum number is the $U(1)_R$ charge.
% The above superconformal primary
% spectrum (\eqref{N2spectrum_1}) arises from the ${\cal N}=1$ spectrum as follows. We consider
% the ${\cal N}=1$ (integer) spin $s$ and spin $s+\half$ superconformal multiplet and arrange their respective conformal
% contents into one (integer) spin $s$ ${\cal N}=2$ superconformal multiplet; it may be checked that this is conserved
% (i.e., short) with respect to the two $D_\alpha^{(i)}$, $i=1,2$ operators. In this way, beginning with
% spin $0$, we see that the whole spectrum of integer currents of the ${\cal N}=2$ theories can be built. Note that
% all currents are $U(1)_R$ singlets.

\subsection{${\cal N}=3$}\label{n3spectrum}
The conformal content of the ${\cal N}=3$ theory is\footnote{For $R$-symmetry quantum numbers taking values in $SU(2)_R$, we give the
dimension of the representation while writing down the quantum numbers $(\Delta,j,h)$. For example, $(1,0,1)$ corresponds
to $\Delta=1$, spin-$0$ and a singlet under $R$. In other words, instead of writing the highest weight $j$ for the
$R$-symmetry representation, we write $2j+1$ as the third quantum number}
\begin{eqnarray}\label{confcont}
   \left[2\bigoplus_{j=0,\half,1,\ldots}^\infty\left[(j+1,j,1) \oplus (j+1,j,3)\right]\right] \oplus (1,0,1) \oplus (1,0,3) \oplus (2,0,1) \oplus (2,0,3)
\end{eqnarray}
The above conformal content can be grouped into ${\cal N}=3$ superconformal primary content as follows
\begin{equation}\label{spectrumN3}
 \left[\bigoplus_{j=0,\half,1,\ldots}^\infty (j+1,j,1)_{{\cal N}=3} \right] \oplus (1,0,3)_{{\cal N}=3}
\end{equation}
The decomposition of the ${\cal N}=3$ superconformal primaries into
${\cal N}=2$ superconformal primaries, given in \cite{Minwalla:2011ma} is\footnote{\label{18}In the equation that follows 
note that the L.H.S. is written in
terms of the $SU(2)_R$ quantum number whereas the R.H.S. has $U(1)_R$ quantum numbers.}
\begin{eqnarray}\label{breakup}
  \begin{array}{ccc}
   (j+1,j,1)_{{\cal N}=3} \longrightarrow & (j+1,j,0)_{{\cal N}=2} ~~\oplus~~ & (j+\frac{3}2,j+\half,0)_{{\cal N}=2}
  \end{array}
\end{eqnarray}
We have the following result
for the conformal content of a ${\cal N}=3$ superconformal primary of spin $j \in (0,\half,1,\ldots)$:
\begin{equation}\label{breakup2}
 (j+1,j,1)_{{\cal N}=3} \rightarrow (j+1,j,1) \oplus (j+\frac{3}2,j+\half,3) \oplus (j+2,j+1,3) \oplus (j+\frac{5}2,j+\frac{3}2,1)
\end{equation}
The breakup of the $(1,0,3)_{{\cal N}=3}$ superconformal primary
into ${\cal N}=2$ primaries is as follows\footnotemark[18]
\begin{eqnarray}\label{breakup3}
 \begin{array}{cc}
   (1,0,3)_{{\cal N}=3} \rightarrow& (1,0,1)_{{\cal N}=2} \oplus (1,0,0)_{{\cal N}=2} \oplus (1,0,-1)_{{\cal N}=2}
 \end{array}
\end{eqnarray}
The conformal content of
the $(1,0,3)_{{\cal N}=3}$ superconformal primary is\footnote{written out in $SU(2)_R$ notation}:
\begin{equation}\label{breakup2}
 (1,0,3)_{{\cal N}=3} \rightarrow (1,0,3) \oplus (2,0,3) \oplus (\frac{3}2,\half,1) \oplus (\frac{3}2,\half,3) \oplus (2,1,1)
\end{equation}

\subsection{${\cal N}=4$}\label{n4spectrum}
The theory contains currents of integer spins only.
It remains to comment about the $(1,0,3)$ superconformal primary which was obtained in equation \ref{spectrumN3}
above. This particular primary transforms in the antisymmetric $(1,0)$ representation of the $SO(4) \sim SU(2) \times SU(2)$
$R$-symmetry, where the two numbers correspond to each of the two $SU(2)$s. Therefore we have
\begin{equation}\label{spectrumN4}
 \left[\bigoplus_{j=0,1,\ldots}^\infty (j+1,j,\{0,0\})_{{\cal N}=4} \right] \oplus (1,0,\{1,0\})_{{\cal N}=4}
\end{equation}
where by $\{0,0\}$ we mean the singlet of the $SU(2) \times SU(2)$ $R$-symmetry.

\subsection{${\cal N}=6$}\label{n6spectrum}
From the field content of the ${\cal N}=6$ theory (see section \ref{supsp6})
the conformal primary spectrum can be easily read off as
\be\begin{split}\label{N6spec}
&      \left[2\bigoplus_{j=0,1,\ldots}^{\infty} \left((j+2,j+1;1) \oplus (j+2,j+1;15) \oplus 
                         (j+\frac{3}{2},j+\half;6) \oplus (j+\frac{3}{2},j+\half;10) \right)\right]\\
&\oplus(1,0;1) \oplus (1,0 ; 15) \oplus (2,0;1) \oplus (2,0;15) \oplus 
\end{split}
\ee
where the conformal primaries are labeled as ($\Delta$\,,\,$j$\,;\,$SO(6)$ representation). 
In specifying the $SO(6)$ $R$-symmetry representation we use the following notation
\be\begin{split}
1& \rightarrow \text{Singlet}, \\ 
6& \rightarrow \text{Vector}, \\ 
15& \rightarrow \text{Second rank symmetric traceless tensor}, \\ 
10& \rightarrow \text{(anti)Self-dual 3 form}. 
\end{split}
\ee

The conformal primary spectrum can be grouped together into the following ${\cal N}=6$ superconformal 
primary spectrum 
\be
 \left[\bigoplus_{j=1,2,\ldots}^\infty (j+1,j;1)_{{\cal N}=6}\right] \oplus (1, 0 ; 15)_{{\cal N}=6}
\ee   
where again use the same labeling for the superconformal primary as above for conformal primaries 
with and extra subscript to distinguish from conformal primaries.

\bibliographystyle{JHEP}
\bibliography{3dscft}

\end{document}